\documentclass[fleqn,usenatbib]{mnras}

\usepackage{newtxtext,newtxmath}

\usepackage[T1]{fontenc}

\DeclareRobustCommand{\VAN}[3]{#2}
\let\VANthebibliography\thebibliography
\def\thebibliography{\DeclareRobustCommand{\VAN}[3]{##3}\VANthebibliography}

\usepackage{graphicx}	
\usepackage{amsmath}	

\usepackage{tikz,xparse}
\usetikzlibrary{positioning}
\usetikzlibrary{arrows.meta,positioning,calc,fit}
\NewDocumentCommand{\linkterms}{ O{1ex} m O{} m m }{
  \tikz[remember picture,baseline=(A.base)]{\node[inner xsep=0pt] (A) {$#2$};}
  #3
  \tikz[remember picture,baseline=(C.base)]{\node[inner xsep=0pt] (C) {$#4$};}
  \tikz[remember picture,overlay]{
    \draw (A.south) -- ([yshift=-#1]A.south) -- coordinate (Z) ([yshift=-#1]C.south) -- (C.south);
    \draw (Z) -- +(0,-#1) node[below] {$#5$};
  }
}

\usepackage{booktabs}
\usepackage{array,tabulary,tabularx,multirow}
\usepackage{tikz,pgfplots}
\pgfplotsset{compat=1.18}
\usepackage{epstopdf}
\usepackage{graphicx}
\usepackage{subcaption}
\usepackage{anyfontsize}
\usepackage{lipsum}
\usetikzlibrary{patterns}

\DeclareMathAlphabet\mathbfcal{OMS}{cmsy}{b}{n}

\newcommand{\justify}[1]{\par\begingroup\parfillskip0pt #1\par\endgroup}

\usepackage[precision=2, unit=mm]{lengthconvert}

\title[Direct Primary Beam Correction]{Direct Primary Beam Correction: A Regularised Linear Operator for Untangling Mutual Coupling in SKA-Low 21-cm Cosmology}

\author[O'Hara et al.]{\parbox{\textwidth}{
Oscar S.D. O'Hara$^{1,2}$,\thanks{E-mail: osdo2@cam.ac.uk}
Quentin Gueuning$^{1,2}$,
Eloy de Lera Acedo$^{1,2}$,
John Cumner$^{1,2}$,
Dominic Anstey$^{1, 2}$,
Anthony Brown$^{3,4}$,
Fred Dulwich$^{1}$,
Andrew Faulkner$^{1}$,
Ashish Mhaske$^{1,2}$,
Oskar Zetterstrom$^{1,2,5}$
\\
\em{\small\vspace*{-0.5\baselineskip}
$^{1}$Cavendish Astrophysics, University of Cambridge, Cambridge, CB3 0HE, UK \\ \vspace*{-0.5\baselineskip}
$^{2}$Kavli Institute for Cosmology in Cambridge, University of Cambridge, Cambridge, CB3 0HA, UK\\ \vspace*{-0.5\baselineskip}
$^{3}$Queen Mary, University of London, London, E1 4NS, UK\\ \vspace*{-0.5\baselineskip}
$^{4}$University of Manchester, Manchester, M13 9PL, UK\\ \vspace*{-0.5\baselineskip}
$^{5}$Department of Communication Systems, KTH Royal Institute of Technology, Stockholm, 11428, Sweden
}}}

\date{Accepted XXX. Received YYY; in original form ZZZ}

\pubyear{\the\year{}}

\begin{document}
\label{firstpage}
\pagerange{\pageref{firstpage}--\pageref{lastpage}}
\maketitle

\begin{abstract} 
\justify{\noindent Mutual coupling has emerged as the dominant direction-dependent corruption in dense aperture arrays, imprinting spatial and sub-MHz spectral structure that compromises time-gating and foreground-removal strategies. We introduce \textit{direct primary beam correction}, a domain-agnostic, regularised, direction-weighted linear inversion of stacked Jones matrices that transforms an aperture array's far-field pattern towards an arbitrary target. Using full-wave electromagnetic simulations of SKA-Low and a 4-hour observation of the EoR0 field spanning 122--134~MHz, we show that the correction reproduces the target pattern exactly for as many directions as there are elements, with regularisation trading in-field accuracy for sidelobe conditioning. For 21-cm power-spectrum analysis, we introduce a cross-time estimator that correlates samples within each $uv$ cell only between different time blocks. As the field is tracked, sidelobe emission decorrelates between blocks, suppressing its coupling-induced leakage into the EoR window; however, the main beam's coupling-induced chromaticity remains coherent and biases the recovered power, a bias that longer integration does not remove. Combined with the correction, the estimator recovers the input power without bias and passes a foreground-only null test, trading the systematic bias for statistical scatter that averages down over independent correction cycles. Propagating Gaussian beam-model errors shows that detection requires embedded element patterns accurate to at least 3.25 significant figures for channel-independent errors. In contrast, common approximations, such as averaged or neighbouring-channel patterns, reach at most two. Mutual coupling is therefore correctable in principle, making beam characterisation, rather than the correction, the priority for 21-cm experiments with dense aperture arrays.}
\end{abstract}




\section{Introduction}

\justify{Over the past decade, increasingly stringent upper limits on the 21-cm power spectrum have begun to place meaningful constraints on theoretical models of the Cosmic Dawn and Epoch of Reionisation (EoR) \citep{skao_book_2026}. As construction of the Square Kilometre Array Low-frequency telescope (SKA-Low; \citealt{labate2022highlights}) progresses toward completion, the instrument is expected to measure spatial fluctuations in the redshifted 21-cm signal with unprecedented sensitivity at metre wavelengths, offering the prospect of directly probing how the first luminous sources transformed the early Universe. Realising this scientific potential requires characterising and mitigating chromatic direction-independent (DI) and direction-dependent (DD) effects. If left unchecked, these effects imprint sharp-scale spectral and spatial structure that drive the nominally band-limited foreground emission to regions of Fourier space (k-space) expected to contain the cosmological 21-cm signal, thereby fundamentally limiting the time-gating and foreground-separation strategies used to isolate the faint 21-cm signal from bright astrophysical foregrounds \citep{kern2019mitigating}.}

\justify{Strong electromagnetic interactions between antennas, commonly referred to as mutual coupling, have been identified over the past year as the dominant DD systematic limiting further progress. Recent Phase II results from the Hydrogen Epoch of Reionization Array (HERA; \citealt{deboer2017hydrogen}) demonstrate that coupling-induced foreground contamination dominates Fourier modes up to $k \sim 0.7\,h\,\mathrm{cMpc}^{-1}$ \citep{TheHERACollaboration2025FirstII}. Despite efforts to suppress this contamination by filtering coupled foreground power in the fringe-rate domain, such techniques have proven inadequate, thereby necessitating more advanced mitigation strategies \citep{rath_2025}. Compounding this challenge, full-wave electromagnetic simulations of SKA-Low reveal strong, sub-MHz spectral structure arising from mutual coupling \citep{davidson_2020, bolli_2022}, which drives foreground leakage beyond $k_{\parallel} \sim 2\,h\,\mathrm{cMpc}^{-1}$ and obscures the 21-cm signal by several orders of magnitude \citep{o2025uncovering}. This situation highlights a fundamental design tension: next-generation instruments require compact interferometric configurations to access the short baselines and Fourier modes most sensitive to the cosmological 21-cm signal, yet such compactness unavoidably leads to significant mutual coupling. Although optimised array layouts have been shown to alleviate some associated negative impacts, such as zenith blindness \citep{anstey_2024}, they do not fully resolve the issue, thereby motivating continued investigation into more effective coupling mitigation methodologies.}

In current 21-cm analysis pipelines, the mitigation of such instrumental systematics rests on two explicit procedures: \textit{calibration} and \textit{correction} \citep{smirnov2011revisitingII}.
\begin{itemize}
    \item \justify{\textit{Calibration} solves for the complex antenna gains, more generally Jones matrices \citep{jones1941new}, that describe how the instrument and propagation path corrupt the incoming signal. Most techniques fit a sky model to the measured visibilities via the radio interferometric measurement equation (RIME; \citealt{smirnov2011revisitingI}), beginning with DI self-calibration \citep{cornwall1981selfcal} and refining with DD Jones terms. Examples include redundant calibration \citep{liu_redundant_2010} for HERA, real-time peeling \citep{mitchell2008rts} for the Murchison Widefield Array (MWA; \citealt{tingay_mwa}), and DDECAL \citep{gan2023assessing} for the Low-Frequency Array (LOFAR; \citealt{van2013lofar}) and NenuFAR \citep{zarka_nenufar}. These schemes depend on accurate sky models, which are themselves degraded by DD effects such as the ionosphere \citep{koopmans2010ionospheric}, and typically assume spectrally smooth gains and identical element patterns across the array. Mutual coupling violates both assumptions: beam-model errors imprint sub-MHz structure onto calibration solutions that smooth, low-order gain models cannot absorb \citep{wijnholds2019using, brackenhoff2025robust}, leaving residuals that can mask or mimic the 21-cm signal. Incorporating each antenna's embedded element pattern (EEP) into calibration improves performance \citep{wijnholds2020embedded}. Still, sub-per cent accuracy across fine frequency channels remains limited by how well EEPs can be characterised, whether through electromagnetic simulation \citep{ha_2018, davidson_2020, bolli_2022, conradie_2023, gueuning_2025}, UAV-based mapping \citep{paonessa_2023}, satellite beacons \citep{chokshi_2021}, or holography \citep{amiri2024holographic}.}

    \vspace{0.1cm}
    \item \justify{\textit{Correction} is the subsequent application of these solutions to the data. Unlike a DI gain, a DD effect cannot be removed by a simple multiplicative factor, because the corrupted visibilities are no longer direct Fourier samples of the sky \citep{smirnov2011revisitingII}. Four approaches predominate. Direct $uv$-plane subtraction forward-models the sky through the RIME, including DD terms, and subtracts the result; it is exact for modelled sources but leaves unmodelled flux uncorrected, a significant limitation given that current all-sky maps still require substantial revision \citep{mkay_sky_model_correction}, and it is computationally expensive at the required fidelity \citep{ohara_modelling}. Facet imaging \citep{tasse2018faceting} divides the field into small sub-images, each corrected using the Jones term at its centre; errors grow away from facet centres, the cost scales with the number of facets, and DD effects that evolve during an observation are poorly handled. AW-projection \citep{bhatnagar2008correcting} instead applies corrections as per-antenna convolution kernels during gridding, but in phased arrays where every element has a unique EEP, each station requires its own kernel and the method scales poorly. Image-domain gridding (IDG; \citealt{van2018image}) avoids explicit kernels altogether by gridding visibilities onto small subgrids and applying DD terms as image-domain multiplications, accommodating station- and time-dependent beams at a cost that maps efficiently onto GPUs.}
\end{itemize}

\justify{The approaches above act on visibilities after correlation. For phased-array instruments such as SKA-Low, however, the station beam is itself formed by complex weights applied to each element in the station signal-processing chain \citep{comoretto2017signal}, so some DD effects can be mitigated before they enter the visibilities. Apodisation tapers the weights across the station aperture to suppress sidelobes, reducing the response to bright off-axis emission at the cost of a broader main lobe and lower sensitivity \citep{taylor1955design, mort2016analysing}. Deterministic nulling goes further, placing nulls in the directions of known bright sources \citep{bui2018direct}. However, nulling alters the array’s response to the sky, and must be tracked or compensated before imaging \citep{leshem2000multichannel}. Where the EEPs are known, weights can also compensate for mutual coupling, either by reconstructing the coupling-free element response \citep{huang2013mutual} or by correcting the element weights through the array coupling matrix \citep{steyskal1990mutual}. Frequency-invariant beamforming, in turn, scales the active aperture with wavelength so that the array factor keeps a constant shape across the band \citep{ward1995theory}, at the cost of using only part of the aperture at higher frequencies; because it assumes identical, uncoupled elements, it cannot remove chromaticity introduced by the EEPs themselves. All of these techniques presuppose accurately calibrated element gains \citep{benthem2021aperture, macario2022characterization}, and since station calibration solves for gains towards a calibrator, antenna-to-antenna EEP variations are absorbed into the solutions unless explicitly modelled \citep{jishnu_2024}.}

\justify{In this work, we present \textit{direct primary beam correction}, a regularised, direction-weighted linear operator that transforms an aperture array's far-field response towards an arbitrary target. Rather than dividing out the beam after imaging, as in conventional primary-beam correction, it reshapes the pattern during beamforming within a chosen field of view so the data carries the target response before any processing. It does not replace direction-dependent calibration, which must still track the ionosphere; instead, it removes the sub-MHz, element-dependent structure that calibration cannot absorb, restoring spectrally smooth gains and identical element patterns within the constrained field of view. Here, we use it to remove mutual coupling by mapping each dense-array EEP onto the isolated element pattern (IEP). Section~\ref{sec:modelling} outlines the mathematical framework for modelling interferometric visibilities, including mutual coupling effects; derives the antenna-level correction; and describes the estimation of the 21-cm power spectrum. Section~\ref{sec:ska_low} presents a simulated SKA-Low case study: we describe the simulation set-up (Section~\ref{sec:sim_setup}), apply the correction with an exact beam model (Section~\ref{sec:exact_model}), assess its effect on EoR window recovery (Section~\ref{sec:mainbeam_sidelobe}) and its robustness to beam-model errors (Section~\ref{sec:robustness}), and derive the beam-model requirements for a 21-cm detection (Section~\ref{sec:requirements}). Section~\ref{sec:conclusions} presents our conclusions.}

\vspace{0.1cm}
\justify{\textit{Notation:} Scalars are denoted by lowercase italics ($a$), vectors by lowercase boldface ($\mathbf{a}$), unit vectors by hats ($\hat{\mathbf{a}}$), and matrices by uppercase boldface ($\mathbf{A}$). The matrix dimensions and domain are indicated by $\in \mathbb{C}^{m \times n}$, where the blackboard bold symbol specifies the set of complex-valued elements and the superscript denotes the matrix size. The operators $(\cdot)^{\ast}$, $(\cdot)^{\mathrm{T}}$, $(\cdot)^{\dagger}$, and $(\cdot)^{+}$ denote the complex conjugate, matrix transpose, Hermitian transpose, and Moore--Penrose pseudoinverse, respectively. Angle brackets $\langle\cdot\rangle$ denote an average over a correlator integration of length $\Delta t$, labelled by its centre time $t$, and $||\cdot||_{\mathrm{F}}$ denotes the Frobenius norm. The row-wise Khatri--Rao and Kronecker products are written as $\odot$ and $\otimes$, respectively, and $\mathbf{I}_n$ denotes the $n \times n$ identity matrix. Matrix vectorisation uses $\mathrm{vec}(\cdot)$, and $\mathrm{diag}(\cdot)$ extracts the diagonal of its argument. The imaginary unit is written $\mathrm{j}$.}

\section{Modelling and Correcting Mutual Coupling in Phased-Array Interferometers}
\label{sec:modelling}

Consider a station composed of $N_a$ dual-polarised antennas. Each antenna provides measurements from two orthogonal feeds, indexed by $p \in \{\mathrm{x}, \mathrm{y}\}$ and aligned with unit vectors $\hat{\mathbf{x}}$ and $\hat{\mathbf{y}}$. The complex voltage measured at antenna $n$ is described by the vector
\begin{equation}
    \mathbf{v}_n(f,t) = 
    \begin{bmatrix} 
    v_{n,\mathrm{x}} \\[3pt] 
    v_{n,\mathrm{y}} 
    \end{bmatrix},
\end{equation}
with components $v_{n,p}(f,t)$ for each polarisation. These voltages are induced by an incident plane wave with an electric field
\begin{equation}
    \mathbf{e}_{\mathrm{i}}(f,\hat{\mathbf{s}},t) = 
    \begin{bmatrix} 
    e_{\mathrm{i},\mathrm{x}} \\[3pt] 
    e_{\mathrm{i},\mathrm{y}} 
    \end{bmatrix},
\end{equation} 
arriving from direction $\hat{\mathbf{s}} = l_x\,\hat{\mathbf{x}} + m_y\,\hat{\mathbf{y}} + n_z\,\hat{\mathbf{z}}$, where $l_x$, $m_y$, and $n_z = (1 - l_x^2 - m_y^2)^{1/2}$ are the direction cosines. Using the RIME formalism outlined in \citet{smirnov2011revisitingI}, the voltage measured at the $n$-th antenna can be related to this incident field through a linear transformation:
\begin{equation}
    \begin{bmatrix} 
    v_{n,\mathrm{x}} \\[3pt] 
    v_{n,\mathrm{y}} 
    \end{bmatrix}
    =     
    \underbrace{\begin{bmatrix}
        J_{n,\mathrm{x}\mathrm{x}} & J_{n,\mathrm{x}\mathrm{y}}\\[3pt]
        J_{n,\mathrm{y}\mathrm{x}} & J_{n,\mathrm{y}\mathrm{y}}
    \end{bmatrix}}_{\mathbf{J}_{n}(f,\hat{\mathbf{s}})}    
    \begin{bmatrix} 
    e_{\mathrm{i},\mathrm{x}} \\[3pt] 
    e_{\mathrm{i},\mathrm{y}} 
    \end{bmatrix},
\label{eq:volt_to_field}
\end{equation}
where $\mathbf{J}_n(f,\hat{\mathbf{s}})$ is the Jones matrix of antenna $n$ \citep{craeye2011review},
\begin{equation}
    \mathbf{J}_n(f,\hat{\mathbf{s}}) = 
    \frac{2\lambda_0}{\mathrm{j} \eta_0}
    \underbrace{
    \begin{bmatrix}
        Z_{n,\mathrm{x}} & 0 \\[3pt]
        0 & Z_{n,\mathrm{y}}
    \end{bmatrix}}_{\mathbf{Z}_{n}(f)}
    \underbrace{
    \begin{bmatrix}
        \mathrm{f}_{n,\mathrm{x}\mathrm{x}} & \mathrm{f}_{n,\mathrm{x}\mathrm{y}}\\[3pt]
        \mathrm{f}_{n,\mathrm{y}\mathrm{x}} & \mathrm{f}_{n,\mathrm{y}\mathrm{y}}
    \end{bmatrix}}_{\mathbf{f}_{n}(f,\hat{\mathbf{s}})}.
\end{equation}
Here, $\eta_0$ and $\lambda_0$ denote the free-space impedance and wavelength, respectively. The matrix $\mathbf{f}_n(f,\hat{\mathbf{s}})$ represents the EEP across both polarisations, with the phase reference taken at the antenna centre. The diagonal matrix $\mathbf{Z}_n(f)$ captures the input impedances of the front-end electronics connected to the respective feeds.

Let $\mathbf{j}_{n,p}=\begin{bmatrix} J_{n,p\mathrm{x}} & J_{n,p\mathrm{y}} \end{bmatrix}$ denote the $p$-th row of $\mathbf{J}_n$, which describes how feed $p$ of antenna $n$ responds to the two polarisation components of the incident field. For each feed $p$, we introduce the array steering matrix $\mathbfcal{J}_p(f,\hat{\mathbf{s}})\in\mathbb{C}^{N_a\times 2}$,
\begin{equation}
    \mathbfcal{J}_p (f,\hat{\mathbf{s}}) =
    \begin{bmatrix}
        \mathbf{j}_{1,p} \, e^{\mathrm{j} k_0 \, \hat{\mathbf{s}} \cdot \mathbf{r}_1}\\[3pt]
        \vdots \\[3pt]
        ~\mathbf{j}_{N_a,p} \, e^{\mathrm{j} k_0 \, \hat{\mathbf{s}} \cdot \mathbf{r}_{N_a}}
   \end{bmatrix},
\end{equation}
where the exponential term accounts for the geometric delay at the antenna position $\mathbf{r}_n$ relative to the station phase reference, and $k_0$ is the free-space wavenumber. The two columns of $\mathbfcal{J}_p$ represent feed $p$'s response to the $\mathrm{x}$- and $\mathrm{y}$-polarised components of the incident field, incorporating both co-polar and cross-polar contributions.

The array pattern is the far-field response of the station, accounting for the position and Jones matrix of each antenna. It is steered towards direction $\hat{\mathbf{s}}_l$ through a coherent weighted sum at the beamformer (Figure~\ref{fig:beamformed}, Beamformers); the corresponding array pattern is given as
\begin{equation}
\begin{split}
    \mathbf{A}_{l}(f,\hat{\mathbf{s}}) &= 
    \begin{bmatrix}
        \mathbf{w}_l & 0 \\
        0 & \mathbf{w}_l
    \end{bmatrix}
    \begin{bmatrix}
        \mathbfcal{J}_\mathrm{x}(f,\hat{\mathbf{s}}) \\[3pt]
        \mathbfcal{J}_\mathrm{y}(f,\hat{\mathbf{s}})
    \end{bmatrix}
\end{split}
\label{eq:array_pattern}
\end{equation}
where $\mathbf{w}_l \in \mathbb{C}^{1 \times N_a}$ contains the complex beamforming weights, with entries
\begin{equation}
    w_{l,n} = e^{-\mathrm{j} k_0 \hat{\mathbf{s}}_l \cdot \mathbf{r}_n}.
\end{equation}
Each weight cancels the geometric phase of antenna $n$ for a wave arriving from $\hat{\mathbf{s}}_l$, so that signals from that direction add in phase. The same weights are applied to both feeds; however, this need not always be the case. 

The complex visibility on the baseline $\mathbf{b}_{mn} = \mathbf{r}_m - \mathbf{r}_n$ is obtained by correlating the beamformed voltages of stations $m$ and $n$, both pointed towards $\hat{\mathbf{s}}_l$ (Figure~\ref{fig:beamformed}, Correlators). For a spatially incoherent sky, emission from different directions is uncorrelated, and the visibility reduces to a single integral over the sky, given by
\begingroup
\small
\begin{equation}
\label{eq:visibility_rime}
    \mathbf{R}_{l,mn}(f,t) = \int \mathbf{A}_{l,m}(f,\hat{\mathbf{s}}) \, \mathbf{I}(f,\hat{\mathbf{s}}, t) \, \mathbf{A}_{l,n}^{\dagger}(f,\hat{\mathbf{s}}) \, e^{\mathrm{j} k_0 \hat{\mathbf{s}} \cdot \mathbf{b}_{mn}} \, \mathrm{d}\Omega.
\end{equation}
\endgroup
Here, the integral runs over the visible hemisphere, $\mathbf{A}_{l,m}$ and $\mathbf{A}_{l,n}$ are the array patterns of the two stations, the exponential term accounts for the geometric delay between them, and $\mathbf{I}(f,\hat{\mathbf{s}},t) = \langle \mathbf{e}_\mathrm{i}\, \mathbf{e}_\mathrm{i}^\dagger \rangle$ is the brightness matrix of the incident field, averaged over the integration centred on $t$. Because $\hat{\mathbf{s}}$ is defined in the station frame, this remaining time dependence captures the sky's motion relative to the array.

\begin{figure}
\centering
\vspace{0.5cm}
\resizebox{1.0\linewidth}{!}{
\begin{tikzpicture}[scale=\textwidth/22cm,samples=200]
    \def\n{5}        
    \def\spacing{0.8}
    \def\rot{20}
    
    \newcommand{\lmgrid}[2]{ 
      \begin{scope}[shift={(#1,#2)}, xscale=1.25, yscale=0.6125, rotate={-45}]
        \foreach \i in {-2,-1,0,1,2} {
          \foreach \j in {-2,-1,0,1,2} {
            \fill (\i*\spacing,\j*\spacing) circle (2.5pt);
          }
        }
        
        \draw[<-,thick] (-0.5*\n*\spacing + .3, 2.2) -- (0.5*\n*\spacing - .3, 2.2);
        \node[above] at (0,2.2) {$l_x$};
    
        \draw[<-,thick] (2.2, -0.5*\n*\spacing + .3) -- (2.2, 0.5*\n*\spacing - .3);
        \node[below] at (2.3,0) {$m_y$};
      \end{scope}
    }
    
    \lmgrid{0}{.7}
    
    \def\beamLength{3}   
    \def\beamWidth{0.35}   
    
    \newcommand{\steering}[4][1]{ 
      \begin{scope}[shift={(#2,#3)}, rotate=#4]
        \draw[black, thick, opacity=#1]
          (0,0)  
            .. controls (-\beamWidth, 0.7*\beamLength) and (-\beamWidth, \beamLength) ..
            (0,\beamLength)  
            .. controls (\beamWidth, \beamLength) and (\beamWidth, 0.7*\beamLength) ..
          cycle;
      \end{scope}
    }
    
    \steering[0.2]{-2.5}{-4.7}{25}
    \steering[0.6]{-2.5}{-4.7}{12.5} 
    \steering[1.0]{-2.5}{-4.7}{0} 
    \steering[0.6]{-2.5}{-4.7}{-12.5} 
    \steering[0.2]{-2.5}{-4.7}{-25}  
    
    \steering[0.2]{2.5}{-4.7}{25}
    \steering[0.6]{2.5}{-4.7}{12.5} 
    \steering[1.0]{2.5}{-4.7}{0} 
    \steering[0.6]{2.5}{-4.7}{-12.5} 
    \steering[0.2]{2.5}{-4.7}{-25}  
    
    \newcommand{\station}[2]{ 
      \begin{scope}[shift={(#1,#2)}]
        \def\centerOffset{3.775}
    
        \fill[black!10] (\centerOffset-3.75,0.7+1.6) ellipse (2cm and 0.5cm);
        
        \draw [black,very thick] (\centerOffset-1.55-1.1,0.8+1.6) -- (\centerOffset-1.55-1.1,1.088+1.6);
        \draw [black,very thick] (\centerOffset-1.55-1.1,1.088+1.6) -- (\centerOffset-1.55-1.1-0.3,1.088+2);
        \draw [black,very thick] (\centerOffset-1.55-1.1,1.088+1.6) -- (\centerOffset-1.55-1.1+0.3,1.088+2);
        
        
        \draw [black,very thick] (\centerOffset-1.55-2*1.1,0.4+1.6) -- (\centerOffset-1.55-2*1.1,0.688+1.6);
        \draw [black,very thick] (\centerOffset-1.55-2*1.1,0.688+1.6) -- (\centerOffset-1.55-2*1.1+0.3,0.688+2);
        \draw [black,very thick] (\centerOffset-1.55-2*1.1,0.688+1.6) -- (\centerOffset-1.55-2*1.1-0.3,0.688+2);    
        
        
        \draw [black,very thick] (\centerOffset-1.55-3*1.1,0.7+1.6) -- (\centerOffset-1.55-3*1.1,0.988+1.6);
        \draw [black,very thick] (\centerOffset-1.55-3*1.1,0.988+1.6) -- (\centerOffset-1.55-3*1.1+0.3,0.988+2);     \draw [black,very thick] (\centerOffset-1.55-3*1.1,0.988+1.6) -- (\centerOffset-1.55-3*1.1-0.3,0.988+2);  
        
        
        \draw [black] (\centerOffset-1.55-1.1,0.3+1.6) -- (\centerOffset-1.55-1.1,-1.25);
        \draw [black] (\centerOffset-1.55-2*1.1,0.2+1.6) -- (\centerOffset-1.55-2*1.1,-1.25);
        \draw [black] (\centerOffset-1.55-3*1.1,0.3+1.6) -- (\centerOffset-1.55-3*1.1,-1.25);

        \node[text width=0.5cm,align=left] at (\centerOffset-1.55-3*1.1-0.1,1.45) {$\mathbf{v}_{1}$};
        \node[text width=0.5cm,align=left] at (\centerOffset-1.55-2*1.1-0.1,1.45) {$\mathbf{v}_{2}$};
        \node[text width=0.5cm,align=left] at (\centerOffset-1.55-1.1-0.1,01.45) {$\mathbf{v}_{3}$};
        
        \draw [black] (\centerOffset-1.55-1.1,-1.25) -- (\centerOffset-1.55-2*1.1,-2.0);
        \draw [black] (\centerOffset-1.55-2*1.1,-1.25) -- (\centerOffset-1.55-2*1.1,-2.0);
        \draw [black] (\centerOffset-1.55-3*1.1,-1.25) -- (\centerOffset-1.55-2*1.1,-2.0);
        
        \draw [draw=black] (\centerOffset-1.95,-2.5) rectangle (\centerOffset-5.6,-0.4);
        
        \foreach \i in {1,2,3} {
          \draw [black,fill=white] (\centerOffset-1.55-\i*1.1,-1.25) circle (0.15cm);
          \draw [black] (\centerOffset-1.55-\i*1.1-0.1061,-1.25+0.1061) -- (\centerOffset-1.55-\i*1.1+0.1061,-1.25-0.1061);
          \draw [black] (\centerOffset-1.55-\i*1.1+0.1061,-1.25+0.1061) -- (\centerOffset-1.55-\i*1.1-0.1061,-1.25-0.1061);
        }
    
        \node[text width=0.5cm,align=left] at (\centerOffset-1.55-3*1.1-0.39,-0.92) {$w_{l,1}$};
        \node[text width=0.5cm,align=left] at (\centerOffset-1.55-2*1.1-0.39,-0.92) {$w_{l,2}$};
        \node[text width=0.5cm,align=left] at (\centerOffset-1.55-1.1-0.39,-0.92) {$w_{l,3}$};
    
        \draw [black,fill=white] (\centerOffset-1.55-2*1.1,-2.0) circle (0.15cm);
        \draw [black] (\centerOffset-1.55-2*1.1-0.15,-2.0) -- (\centerOffset-1.55-2*1.1+0.15,-2.0);
        \draw [black] (\centerOffset-1.55-2*1.1,-2.0-0.15) -- (\centerOffset-1.55-2*1.1,-2.0+0.15);
        
      \end{scope}
    }
    
    \draw [black] (-2.525,-9.5) -- (0,-14);
    \draw [black] (2.525,-9.5) -- (0,-14);
    
    \station{-2.5}{-7.5} 
    \station{2.5}{-7.5}  
    
    \begin{scope}[shift={(6,-5.2)}, xscale=1, yscale=0.25, rotate={180-45}]
      \def\centerOffset{3.775}
    \end{scope}
    
    \draw [draw=black,fill=white] (0.505-0.2,-13.1-0.2) rectangle (0.505+0.2,-13.1+0.2);
    \node[text width=0.25cm,align=left] at (0.555,-13.1) {$\dagger$};
    
    \draw[->,black] (0, -14) -- (0, -16.15);
    
    \draw [black,fill=white] (0,-14) circle (0.15cm);
    \draw [black] (-0.1061, 0.1061-14) -- (+0.1061,-0.1061-14);
    \draw [black] (0.1061, 0.1061-14) -- (-0.1061,-0.1061-14);
    
    \draw [draw=black,fill=white] (-0.3,-14.75-0.3) rectangle (0.3,-14.75+0.3);
    \node[text width=0.25cm,align=left] at (0.05,-14.75) {$\int$};
    
    \draw [draw=black] (-1.35,-14-1.4) rectangle (1.35,-14+1.5);

    \draw [black,dotted] (-6.0,-4) -- (5.75,-4);
    \draw [black,dotted] (-6.0,-7.5) -- (5.75,-7.5);
    \draw [black,dotted] (-6.0,-11.9) -- (5.75,-11.9);
    
    \node[text width=2.5cm,align=left,rotate=90] at (-5.5,-0.5) {BEAM POINTINGS};
    \node[text width=2.0cm,align=left,rotate=90] at (-5.5,-5.5) {RF SYSTEM};
    \node[text width=2.5cm,align=left,rotate=90] at (-5.5,-9.5) {BEAMFORMERS};
    \node[text width=2.5cm,align=left,rotate=90] at (-5.5,-14) {CORRELATORS};
    \node[text width=2.5cm,align=center,rotate=0] at (-2.5,-1.4) {STATION $m$};
    \node[text width=2.5cm,align=center,rotate=0] at (2.5,-1.4) {STATION $n$};
    
    
    \node[text width=0.5cm,align=left] at (3.1,-9.7) {$b_{l,n}$};
    \node[text width=0.5cm,align=left] at (-3.9,-9.7) {$b_{l,m}$};
    
    \draw [draw=black,dashed,rounded corners,fill=white] (-3.943, -7.05) rectangle (-1,-6.35);
    \node at (-2.4715,-6.7) {\scriptsize $\mathbf{C}_\mathbfcal{J} \, \mathbfcal{J}$};
    \draw [draw=black,dashed,rounded corners,fill=white] (1, -7.05) rectangle (3.943,-6.35);
    \node at (2.4715,-6.7) {\scriptsize $\mathbf{C}_\mathbfcal{J} \, \mathbfcal{J}$};

    \draw [draw=black,dashed,rounded corners,fill=white] (-3.0,-11.35) rectangle (-.5,-10.65);
    \draw [draw=black,dashed,rounded corners,fill=white] (3.0,-11.35) rectangle (.5,-10.65);
    \node at (1.75,-11) {\scriptsize $\mathbf{C}_\mathrm{A} \, \mathbf{A}$};
    \node at (-1.75,-11) {\scriptsize $\mathbf{C}_\mathrm{A} \, \mathbf{A}$};
    
    \draw [draw=black,dashed,rounded corners] (-1.543-0.5,-16.429-0.45) rectangle (1.543+0.5,-16.429+0.45);
    \node at (0,-16.429) {\scriptsize $\mathbf{C_A \, R_{mn}\, C_A}^\dagger$};
\end{tikzpicture}}
\caption{A block diagram of a two-station phased-array interferometer. Within each station, the antenna voltages $\mathbf{v}_n$ are weighted by $w_{l,n}$ and summed to form beams towards directions $\hat{\mathbf{s}}_l$. The beamformed voltages $b_{l,m}$ and $b_{l,n}$ of stations $m$ and $n$ are then correlated and time-averaged to give the visibility $R_{l,mn}$. The dashed boxes mark the stages at which a linear correction to the beam can be applied: to the antenna voltages ($\mathbf{C}_\mathbfcal{J}$), to the beamformed voltages ($\mathbf{C}_\mathrm{A}$), or to the visibilities through $\mathbf{C}_\mathrm{A}\mathbf{R}_{mn}\mathbf{C}_\mathrm{A}^\dagger$. These operators are derived in Section~\ref{sec:element-level-cal} and Appendix~\ref{app:station_vis}; the simulations in this work apply the correction to the antenna voltages.}
\label{fig:beamformed}
\end{figure}

\subsection{Antenna-Level Pattern Correction}
\label{sec:element-level-cal}

\justify{Compensating mutual coupling through a linear operation on the element signals in the receiving chain dates back to \citet{steyskal1990mutual}. Using a uniform linear array, they showed that, for single-mode elements, the inverse coupling matrix restores the isolated-element response in all directions. To extract the coupling coefficients, they assumed that every element sees the same environment, as in the interior of a large regular array. Next-generation arrays such as SKA-Low, however, break both assumptions. Their complex antenna geometries support many current modes \citep{kyriakou_cma, cumner2024effects}, so no direction-independent matrix can restore the isolated response exactly. Their element configurations are also pseudo-random \citep{anstey_2024}, and construction tolerances and element degradation add further differences, so no two elements share the same environment.
The Element Pattern Reconstruction of \citet{huang2013mutual}, developed for direction-of-arrival estimation, relaxes both assumptions by fitting a linear correction that maps the embedded- onto the isolated element patterns for a chosen set of directions. We generalise this mapping using three versions of this far-field response: the \emph{true response}, $(\cdot)^{\mathrm{true}}$, which the instrument actually has but we never know exactly; the \emph{model response}, $(\cdot)^{\mathrm{mod}}$, our best approximation to it; and the \emph{target response}, $(\cdot)^{\mathrm{tgt}}$, the response we would like the instrument to have. We adapt this approach to radio astronomy, deriving a regularised, direction-weighted correction at the antenna level for an arbitrary dual-polarised target response, and showing how it carries through the analogue signal chain (Figure~\ref{fig:beamformed}).}

An aperture, whether a station or a dish, of maximum extent $b_{\max}$ (for a station, the largest separation between two of its elements), is sensitive only to spatial frequencies on the sky of up to $b_{\max}/\lambda$. The sky as seen by the instrument is therefore band-limited and, by the Nyquist--Shannon sampling theorem, can be fully described by samples spaced by $\delta = \lambda/(2 b_{\max})$ in each direction cosine. Because the visible sky is also bounded, the instrument can access only an effectively finite number of independent spatial modes, and any discretisation satisfying this criterion preserves all measurable degrees of freedom.
We begin by selecting $M$ directions $\hat{\mathbf{s}}_m$ with spherical coordinates $(\theta_m,\phi_m)$, spaced no larger than $\delta$ apart at the observing frequency. The steering matrix $\mathbfcal{J}_p(f,\hat{\mathbf{s}})$ is then constructed for each feed $p$ in every direction $\hat{\mathbf{s}}_m$. Let $\mathbfcal{J}_{pq}(f,\hat{\mathbf{s}}) \in \mathbb{C}^{N_a}$ denote the column of $\mathbfcal{J}_p$ giving feed $p$'s response to the $q$-polarised field component. Concatenating this column over the $M$ directions gives
\begin{equation}
    \mathbfcal{J}_{pq}(f) =
    \begin{bmatrix}
        \mathbfcal{J}_{pq}(f,\hat{\mathbf{s}}_1) & \cdots & \mathbfcal{J}_{pq}(f,\hat{\mathbf{s}}_M)
    \end{bmatrix}\in\mathbb{C}^{N_a\times M},
\end{equation}
and stacking these blocks over both feeds and both field components gives the full dual-polarised response
\begin{equation}
    \mathbfcal{J}(f) =
    \begin{bmatrix}
        \mathbfcal{J}_{\mathrm{xx}}(f) & \mathbfcal{J}_{\mathrm{xy}}(f)\\
        \mathbfcal{J}_{\mathrm{yx}}(f) & \mathbfcal{J}_{\mathrm{yy}}(f)
    \end{bmatrix}\in\mathbb{C}^{2N_a\times 2M}.
\end{equation}
At each frequency, the relationship between a target $\mathbfcal{J}^{\mathrm{tgt}}$ and model $\mathbfcal{J}^{\mathrm{mod}}$ patterns may be expressed as a linear transformation of the form, 
\begin{equation}
    \mathbfcal{J}^{\mathrm{tgt}} \simeq \mathbf{C}_\mathbfcal{J}\, \mathbfcal{J}^{\mathrm{mod}},
\end{equation}
where $\mathbf{C}_{\mathbfcal{J}} \in \mathbb{C}^{2N_a \times 2N_a}$ is the correction operator. Here, the correction acts on the $2N_a$ antenna voltages, ordered as the rows of $\mathbfcal{J}$ (all $x$-feeds followed by all $y$-feeds); the diagonal blocks act within each polarisation, and the off-diagonal blocks mix the two polarisations, correcting both cross- and co-polar structure. In the specific case where $\mathbfcal{J}^{\mathrm{tgt}}$ corresponds to the isolated element pattern and $\mathbfcal{J}^{\mathrm{mod}}$ the embedded element pattern, $\mathbf{C}_{\mathbfcal{J}}$ accounts for the structure originating from mutual coupling interaction among each antenna element. For single-mode elements, this is exact and $\mathbf{C}_{\mathbfcal{J}}$ reduces to the inverse coupling matrix of \citet{steyskal1990mutual}.

In general, no exact correction exists, so we estimate $\mathbf{C}_{\mathbfcal{J}}$ by minimising the weighted Frobenius norm
\begin{equation}
    \mathcal{L}(\mathbf{C}_{\mathbfcal{J}}) = \left\| \left( \mathbf{C}_{\mathbfcal{J}} \mathbfcal{J}^{\mathrm{mod}} - \mathbfcal{J}^{\mathrm{tgt}} \right) \mathbf{D} \right\|_{\mathrm{F}}^2,
\end{equation}
where $\mathbf{D} \in \mathbb{R}^{2M \times 2M}$ is a diagonal matrix of non-negative weights, one per sampled direction and field component. Unlike the beamforming weights, these act only in the fit, setting how closely the corrected response must match the target in each direction; a zero weight leaves that direction unconstrained. In this work, the weights are binary, with the direction centred on the pointing set to unity (Section~\ref{sec:mc_ant_volt_cal}). Minimising the cost function $\mathcal{L}$ gives the closed-form weighted least-squares solution
\begin{equation}
    \label{eq:element_ls_solution}
    \mathbf{C}_\mathrm{\mathcal{J}} = \mathbfcal{J}^{\mathrm{tgt}} \mathbf{D} \left( \mathbfcal{J}^{\mathrm{mod}} \mathbf{D} \right)^{+},
\end{equation}
where $(\cdot)^{+}$ denotes the Moore--Penrose pseudoinverse, computed via singular value decomposition. When the number of directions with non-zero weight does not exceed $N_a$, and the array responses in those directions are linearly independent, the target is reproduced exactly in those directions and $\mathbf{C}_{\mathbfcal{J}}$ is the minimum-norm solution; otherwise, it is the least-squares fit. The fit is exact over all directions only if a direction-independent matrix relates the model and target. In the single-mode case above, this holds, and, provided enough directions are weighted for $\mathbfcal{J}^{\mathrm{mod}}\mathbf{D}$ to have full row rank, Equation~\eqref{eq:element_ls_solution} recovers the inverse coupling matrix independently of $\mathbf{D}$. In general, it does not hold, and $\mathbf{D}$ determines where the residual is minimised.

In either case, the inversion can become ill-conditioned when sky coverage is incomplete, or the element patterns are highly correlated. We therefore introduce a Tikhonov regularisation, adding a penalty $\lambda_{\mathrm{reg}} \|\mathbf{C}_{\mathbfcal{J}}\|_{\mathrm{F}}^2$ to $\mathcal{L}$, which gives
\begin{equation}
    \mathbf{C}_{\mathbfcal{J}} = \mathbfcal{J}^{\mathrm{tgt}} \mathbf{D} \left( \mathbfcal{J}^{\mathrm{mod}} \mathbf{D} \right)^\dagger \left[\left( \mathbfcal{J}^{\mathrm{mod}} \mathbf{D} \right) \left( \mathbfcal{J}^{\mathrm{mod}} \mathbf{D} \right)^\dagger + \lambda_{\mathrm{reg}} \mathbf{I}_{2N_a}\right]^{-1},
    \label{eq:regularised}
\end{equation}
where $\lambda_{\mathrm{reg}}$ sets the strength of the regularisation and Equation~\eqref{eq:element_ls_solution} is recovered as $\lambda_{\mathrm{reg}} \to 0$. The penalty keeps $\mathbf{C}_{\mathbfcal{J}}$ from growing large along combinations of antenna voltages that are poorly constrained by the sampled model response, which would otherwise amplify receiver noise and model errors. 

Finally, because the correction is a direction-independent linear operation, and beamforming and correlation are respectively linear and quadratic in the antenna voltages, we can apply an analogous correction after beamforming and correlation (Figure~\ref{fig:beamformed}, Beamformers and Correlators). Appendix~\ref{app:station_vis} fits the array patterns of a single station, stacks them over all beams and feeds into $\mathbf{A}$, and derives a station-level operator $\mathbf{C}_\mathrm{A}$ that acts linearly on the beamformed voltages. It enters the visibilities through the congruence $\mathbf{C}_\mathrm{A}\mathbf{R}_{mn}\mathbf{C}_\mathrm{A}^\dagger$, where $\mathbf{R}_{mn}$ is the matrix of correlations between every beam and feed of station $m$ and every beam and feed of station $n$, which contains the visibilities $R_{l,mn}$.

\subsection{Estimating the 21-cm Power Spectrum}

\justify{Consider the visibilities of Equation~\eqref{eq:visibility_rime} for an observation tracking a fixed phase centre. We index each baseline and time sample, a point in the $uv$-plane, by $s \equiv (mn, t)$, and write $R_{pq}(s,f)$ for the correlation between feed $p$ of station $m$ and feed $q$ of station $n$, converted from flux density to brightness temperature using the Rayleigh--Jeans relation. Through the Fourier kernel in Equation~\eqref{eq:visibility_rime}, each visibility samples a transverse Fourier mode of the beam-weighted sky brightness temperature, while the delay transform approximates a Fourier transform along the line of sight. The squared modulus of the delay-transformed visibilities therefore estimates the power spectrum of the brightness-temperature fluctuations. A baseline of length $|\mathbf{b}_{mn}|$ maps to the transverse wavenumber $k_\perp = 2\pi|\mathbf{b}_{mn}|/(\lambda_\mathrm{ref} X)$, and a delay $\tau$ maps to the line-of-sight wavenumber $k_\parallel = 2\pi\tau/Y$ \citep{morales2004toward}. Here, $\lambda_\mathrm{ref}$ is the wavelength at the centre of the band, corresponding to the redshift $1 + z = f_{21}\lambda_\mathrm{ref}/c$, where $f_{21} \approx 1420.4$~MHz is the rest frequency of the 21-cm line and $c$ is the speed of light. The factors $X = D_\mathrm{M}(z)$ and $Y = c(1+z)^2/[H(z)\,f_{21}]$ convert angular and spectral intervals into transverse and line-of-sight comoving distances \citep{lanman_2020}, where $D_\mathrm{M}$ is the transverse comoving distance \citep{hogg1999distance} and $H(z)$ is the Hubble parameter, both evaluated for the cosmological parameters of \citet{planck_2018}.}

The delay-transformed visibility is
\begin{equation}
\label{eq:delay_transform}
    \tilde{R}_{pq}(s,\tau) = \int \mathcal{W}(f)\, R_{pq}(s,f)\, e^{-2\pi\mathrm{j} f\tau}\, \mathrm{d}f,
\end{equation}
where $\mathcal{W}(f)$ is a self-convolved Blackman--Harris taper that suppresses the spectral leakage caused by the finite bandwidth of the observation. The simulated power spectra in this paper will be reported using the Stokes~I parameter formed as $\tilde{R}(s,\tau) = [\tilde{R}_\mathrm{xx}(s,\tau) + \tilde{R}_\mathrm{yy}(s,\tau)]/2$. 

Following \citet{thyagarajan2025comparison}, the per-baseline estimator averages the squared modulus of the delay-transformed visibilities over the set $\mathscr{S}_\beta$ of samples whose baseline length falls in bin $\beta$,
\begin{equation}
    \tilde{\mathcal{P}}(\beta, \tau) = \frac{X^2Y}{\Omega_\mathcal{A} B_\mathrm{eff}}\, \frac{1}{|\mathscr{S}_\beta|} \sum_{s\in\mathscr{S}_\beta} \left|\tilde{R}(s,\tau)\right|^2,
    \label{eq:pk_perbaseline}
\end{equation}
where $\Omega_\mathcal{A} B_\mathrm{eff}$ is the product of the effective beam solid angle and effective bandwidth, averaged over the observation \citep{parsons2014new}. Each sample in Equation~\eqref{eq:pk_perbaseline} is multiplied by its complex conjugate; therefore, the estimator's expectation includes the power of every component in the data, not just the sky. Thermal noise and any time-variable leakage therefore add a positive bias. 

\justify{To remove this bias, we adapt the time-interleaving approach of \citet{TheHERACollaboration2025FirstII}, which cross-multiplies interleaved integrations within redundant baseline groups so that noise, being independent between them, contributes no bias. We change both groupings. A non-redundant array such as SKA-Low has no groups of identical baselines, so we group samples into $uv$ cells instead. In a tracked observation, moreover, the sky rotates relative to the aperture array and the fringe phase of off-axis emission varies across each cell, so emission decorrelates over time at a rate that increases with angular distance from the pointing. Interleaved integrations are too closely spaced for this decorrelation to occur, so we cross-multiply contiguous time blocks instead. We refer to the result as the cross-time estimator.}

Each $uv$ cell, indexed by $\kappa$, has width $b_{\mathrm{max}}/(2\lambda_\mathrm{ref})$, where $b_{\mathrm{max}}$ is as defined in Section~\ref{sec:element-level-cal}, so that all samples within it probe approximately the same Fourier mode of the sky. The time blocks $\mathcal{T}_i$ are contiguous, and in this work each coincides with one correction interval (Section~\ref{sec:ska_low}). Summing the delay-transformed visibilities of all samples in cell $\kappa$ and block $i$ coherently gives $S_{\kappa,i}(\tau)$, and the estimator is built only from products between different blocks,
\begin{equation}
    \tilde{\mathcal{P}}^{\times t}(\beta, \tau) = \frac{X^2Y}{\sum_{\kappa\in\beta} \Gamma_\kappa}\, \sum_{\kappa\in\beta} \sum_{i\neq j} \Re\left\{ S_{\kappa,i}(\tau)\, S^{\ast}_{\kappa,j}(\tau) \right\},
    \label{eq:pk_crosstime}
\end{equation}
\justify{where the sums run over cells whose $uv$ distance falls in bin $\beta$, and $\Gamma_\kappa$ (Appendix~\ref{app:gamma}) is the expected response of these products per unit sky power. Thermal noise, independent between blocks, therefore adds variance to Equation~\eqref{eq:pk_crosstime} but no bias, as does far-sidelobe emission, to the extent that it decorrelates. Emission near the pointing remains coherent and is retained, together with any spectral structure the slowly varying main lobe imprints. We retain components that remain coherent across blocks: the 21-cm signal, time-stable foregrounds, and instrument-fixed leakage. For both estimators, we map each $(\beta, \tau)$ bin to $(k_\perp, k_\parallel)$ and average in spherical shells of $k = (k_\perp^2 + k_\parallel^2)^{1/2}$. We report the result as the dimensionless power spectrum $\Delta^2(k) = k^3\tilde{\mathcal{P}}(k)/(2\pi^2)$.}

\section{Simulated Case Study: The Square Kilometre Array Low Frequency Telescope}
\label{sec:ska_low}

\justify{To quantify how mutual coupling affects 21-cm power-spectrum measurements, and what correcting it requires, we simulate observations with SKA-Low, whose densely packed stations exhibit strong coupling-induced beam distortions across the 21-cm band \citep{ohara_impact}. In this body of work, we derive the element and array far-field response solely from full-wave electromagnetic simulations, ensuring the instrument's true response is known exactly. In reality, these model patterns could also arise from direction-dependent gain solutions or measurements from holography \citep{kiefner2021holographic} or UAV \citep{paonessa_2023}, if sufficiently resolved. As the true response is known exactly, we likewise exclude thermal noise, ionospheric phase screens, and calibration errors, so any departure from the target response arises directly from the coupling, correction, or errors we deliberately introduce into the model. This lets us address three questions that real data cannot separate: what a correction can recover when the beam is known exactly, and where its limits lie (Sections~\ref{sec:mc_ant_volt_cal} and~\ref{sec:spectral}); whether the leakage introduced by mutual coupling survives estimators that rely on decorrelation in time, and how the correction changes this (Section~\ref{sec:mainbeam_sidelobe}); and how accurately the beam must be known for the correction to support a 21-cm detection (Sections~\ref{sec:robustness} and~\ref{sec:requirements}). The telescope model and the electromagnetic and visibility simulations follow \citet{ohara_modelling}; for further details, such as computational scaling and the beam evaluation strategy, we refer the reader there.}

\subsection{Simulation Set-up}
\label{sec:sim_setup}

\begin{table}
\centering
\caption{Common properties of the simulated SKA-Low observations and the correction applied to them.}
\begin{tabularx}{\columnwidth}{@{}l>{\centering\arraybackslash}X@{}}
\toprule
\multicolumn{2}{c}{\textbf{Visibility Simulation}} \\
\midrule
\midrule \vspace{1mm}
Telescope Layout             & SKA-Low, 512 stations \\ \vspace{1mm}
Station Layout               & Perturbed Vogel, 256 SKALA4 \\ \vspace{1mm}
Element Response             & IEP, EEP, Corrected EEP \\ \vspace{1mm}
Baselines                    & 130\,816 (max.\ $\approx 74$~km) \\ \vspace{1mm}
Frequency Range              & 122--134~MHz \\ \vspace{1mm}
Channel Width                & 100~kHz \\ \vspace{1mm}
Tracked Field                & EoR0, $(0\,\mathrm{h}, -30\degr)$ \\ \vspace{1mm}
Observation Length           & 4~hr \\ \vspace{1mm}
Integration Time             & 10~s \\ 
                             & GLEAM, \\
Sky Model                    & GSM ($5\degr$ radius), \\
                             & \texttt{21cmFAST} ($10\degr \times 10\degr$) \\
\bottomrule
\\
\end{tabularx}
\begin{tabularx}{\columnwidth}{@{}l>{\centering\arraybackslash}X@{}}
\toprule
\multicolumn{2}{c}{\textbf{Direct Primary Beam Correction}} \\
\midrule
\midrule \vspace{1mm}
Applied To                   & Antenna voltages \\ \vspace{1mm}
Fit Directions               & Equal-area Fibonacci lattice \\ \vspace{1mm}
Direction Spacing            & $\lambda/(2b_{\max})$, $b_{\mathrm{max}}\approx38~$m \\ \vspace{1mm}
Directional Weights          & Binary, $16.6\degr$--$15.1\degr$ \\ \vspace{1mm}
Regularisation               & $\lambda_{\mathrm{reg}} = 10^{-4}$ \\ \vspace{1mm}
Update Interval              & 10~min (24 cycles; 12 jackknife groups) \\ 
\bottomrule
\end{tabularx}
\label{tab:sim_params}
\end{table}

SKA-Low is situated within Inyarrimanha Ilgari Bundara in Western Australia and will consist of 512 aperture-array stations. A compact core of 224 stations lies within a diameter of $1\,\mathrm{km}$, while the remaining stations are grouped into clusters of six and distributed at logarithmically increasing radii along three spiral arms, yielding a maximum baseline of approximately $74\,\mathrm{km}$. Each station comprises 256 dual-polarised, log-periodic SKALA4 antenna elements \citep{acedo_skala4}, operating over a 7:1 fractional bandwidth spanning $50$--$350\,\mathrm{MHz}$. Within each station, the elements are distributed over a circular aperture of radius $19\,\mathrm{m}$ with a minimum inter-element spacing of $1.7\,\mathrm{m}$. Rather than a regular grid, the element positions follow the Perturbed Vogel configuration \citep{anstey_2024}, whose deliberate aperiodicity suppresses the zenith nulls that mutual coupling would otherwise induce \citep{cumner_dipoles}. Because the elements are so densely packed, we cannot treat their radiation patterns in isolation, and capturing the resulting coupling requires full-wave modelling of the complete station. Using a wire-mesh model of the SKALA4 element within a Perturbed Vogel station, we computed the IEP and EEPs between $122$--$134\,\mathrm{MHz}$ at a channel resolution of $100\,\mathrm{kHz}$ with the Fast Antenna Simulation Tool \citep[FAST;][]{gueuning_2025}, a Method-of-Moments solver tailored to large, densely packed arrays of geometrically complex elements. These patterns form the far-field beam models supplied to the visibility simulations.

Interferometric visibilities were computed using \texttt{OSKAR} \citep{oskar}, a GPU-accelerated radio-interferometer simulator developed for SKA applications. \texttt{OSKAR} implements the RIME \citep{smirnov2011revisitingI}, evaluating the complex visibility of Equation~\eqref{eq:visibility_rime} as a discrete sum over all catalogued sources with a full treatment of direction-dependent instrumental effects. Each simulation requires two inputs: a telescope model, specifying the station and element positions together with their far-field patterns, as described above, and a sky model, specifying the astrophysical emission to be observed.

Following \citet{bonaldi_2025}, the sky model comprises three components, each simulated independently in \texttt{OSKAR} as a separate measurement set so that any combination can be formed by summing visibilities without re-simulation. Extragalactic point sources are drawn from the GLEAM catalogue \citep{hurley_gleam}. Diffuse Galactic emission is taken from the Global Sky Model \citep[GSM;][]{zheng_gsm} and pixelised into unresolved point sources, as the discrete evaluation of the RIME in \texttt{OSKAR} requires; since Nyquist-sampling the full-sky GSM at the resolution of SKA-Low's longest baselines is computationally prohibitive, we restrict it to a $5\degr$ radius around the tracked field, retaining the emission that couples most strongly to the primary beam. The cosmological 21-cm signal is generated with \texttt{21cmFAST} \citep{mesinger_21cmfast} over a $10\degr \times 10\degr$ field centred on the phase centre. 

Each simulation tracks the EoR0 field, centred on $(\mathrm{RA}, \mathrm{Dec.}) = (0\,\mathrm{h}, -30\degr)$, for 4~hr over 122--134~MHz, in 121 channels of 100~kHz matching the resolution of the FAST patterns, with an integration time of $\Delta t = 10$~s. We include all 512 stations, giving 130\,816 baselines per integration. For the corrected response, we apply the antenna-level correction $\mathbf{C}_\mathbfcal{J}$ of Section~\ref{sec:element-level-cal}, re-deriving it every 10~min with $\mathbf{D}$ centred on the phase centre at the midpoint of each interval (Section~\ref{sec:mc_ant_volt_cal}), giving 24 correction cycles over the observation. The field drifts by about $13\degr$ per hour relative to the station, so within an interval it moves at most ${\sim}1\degr$ from the cap centre, well inside the corrected field of view. Each correction cycle also defines one time block of the cross-time estimator of Equation~\eqref{eq:pk_crosstime}. Uncertainties on the power spectra are estimated by a delete-one-group jackknife over 12 groups of two consecutive cycles each. Table~\ref{tab:sim_params} summarises the simulation and correction parameters.

\subsection{Direct Primary Beam Correction with an Exact Beam Model}
\label{sec:exact_model}

To characterise the correction and assess its effectiveness, we assume that the model matches the true response exactly, $\mathbfcal{J}^{\mathrm{mod}} = \mathbfcal{J}^{\mathrm{true}}$, as we also do in the visibility simulations of Section~\ref{sec:mainbeam_sidelobe}, so that any residual arises from the correction rather than from model mis-specification. First, we choose the direction sampling and regularisation of the correction (Section~\ref{sec:mc_ant_volt_cal}) and examine how it changes the station beam's spectral structure (Section~\ref{sec:spectral}), providing a proxy for the expected extent of foreground contamination into the EoR window.

\subsubsection{Parametrising the Correction}
\label{sec:mc_ant_volt_cal}

\begin{figure}
\centering
\includegraphics[width=.9\linewidth]{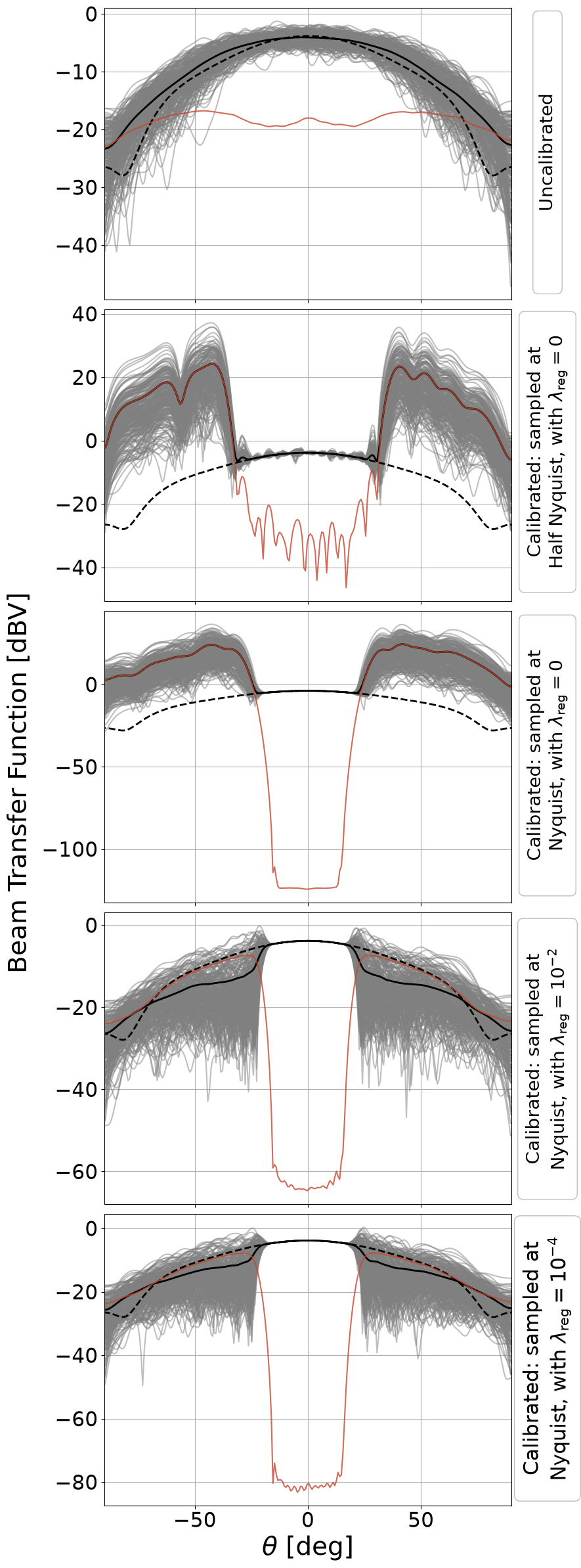}
\caption{Beam transfer function illustrating antenna-level mutual coupling correction applied across a variety of sampling parametrisations and degrees of regularisation at 122~MHz for the perturbed Vogel SKA-Low station. Each panel shows a co-azimuthal cut at $\phi = 0^\circ$ of the Y-polarisation of all embedded element patterns (grey), the corresponding average embedded pattern (solid black), the isolated element pattern (dashed black), and the residual between them (red).}
\label{fig:EP_slices_sampling_and_regularisation}
\end{figure}

\begin{figure*}
    \begin{subfigure}{\linewidth}
    \includegraphics[width=1\linewidth]{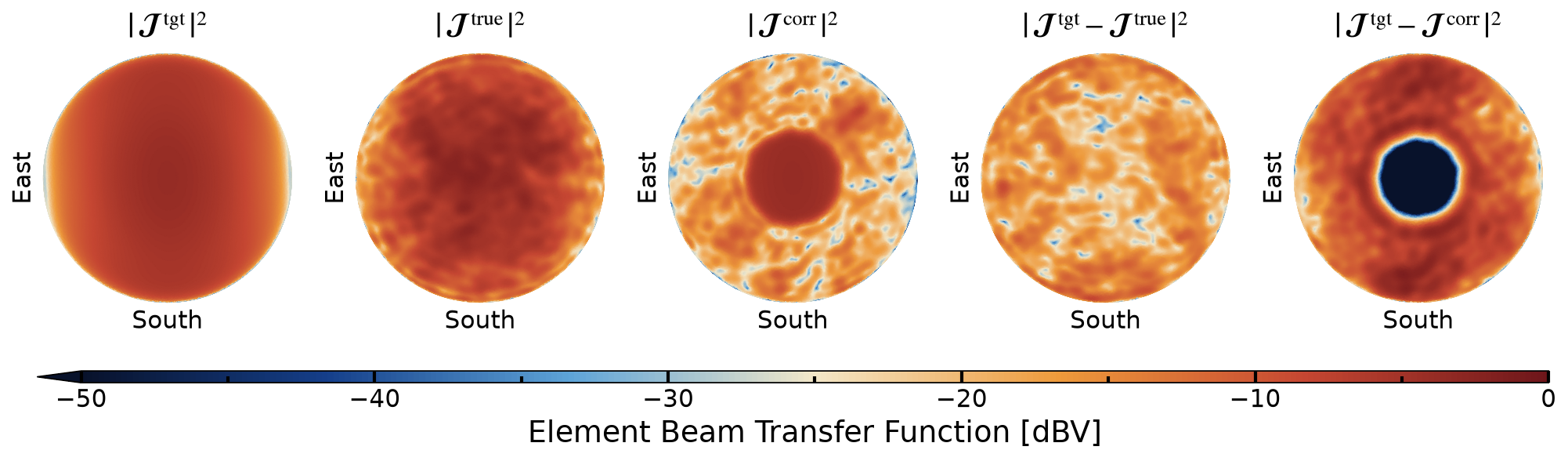}
    \end{subfigure}
    
    \vspace{1.25cm}
    
    \begin{subfigure}{\linewidth}
    \includegraphics[width=1\linewidth]{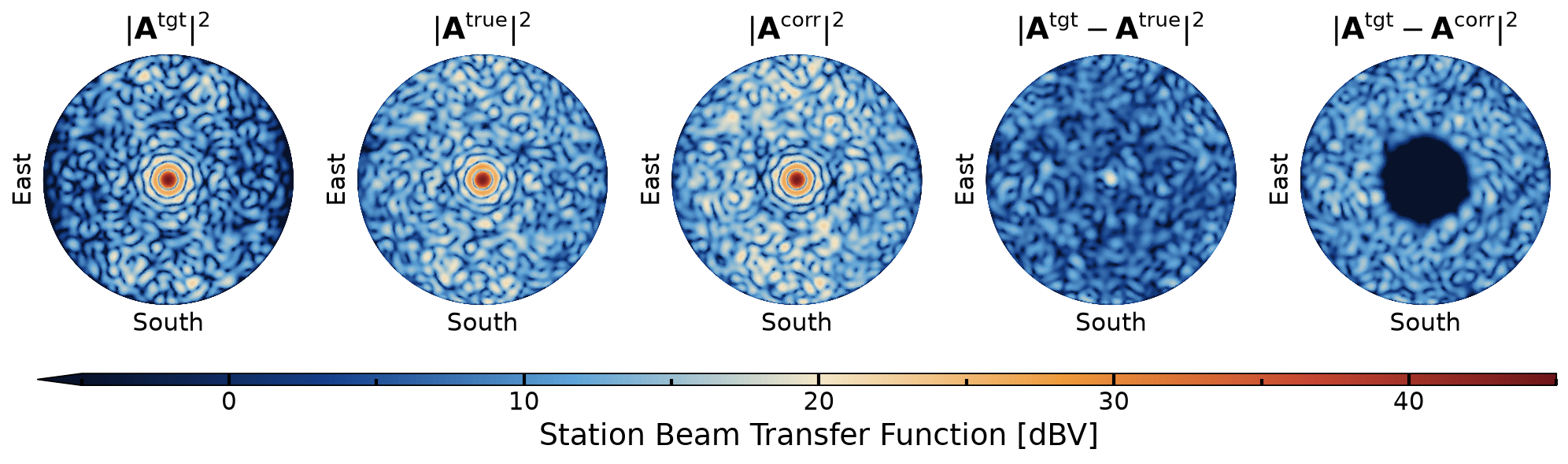}
    \end{subfigure}
    \caption{Illustration of antenna-level mutual coupling correction for a perturbed Vogel SKA-Low station composed of SKALA4 elements at 122~MHz. The top row shows the central element response, and the bottom row shows the resulting array pattern with uniform beamforming weights. From left to right, the panels show the target response, formed from the isolated element pattern ($|\mathbf{f}^{\mathrm{tgt}}_n|$, $|\mathbf{A}^{\mathrm{tgt}}|$); the true response, formed from the embedded element patterns ($|\mathbf{f}^{\mathrm{true}}_n|$, $|\mathbf{A}^{\mathrm{true}}|$); the corrected response after applying $\mathbf{C}_\mathbfcal{J}$ ($|\mathbf{f}^{\mathrm{corr}}_n|$, $|\mathbf{A}^{\mathrm{corr}}|$); and the residuals with respect to the target before correction ($|\mathbf{f}^{\mathrm{tgt}}_n - \mathbf{f}^{\mathrm{true}}_n|$, $|\mathbf{A}^{\mathrm{tgt}} - \mathbf{A}^{\mathrm{true}}|$) and after it ($|\mathbf{f}^{\mathrm{tgt}}_n - \mathbf{f}^{\mathrm{corr}}_n|$, $|\mathbf{A}^{\mathrm{tgt}} - \mathbf{A}^{\mathrm{corr}}|$). All patterns are shown as beam transfer functions in dBV for the Y-polarisation in direction-cosine coordinates. The correction employs Tikhonov regularisation with $\lambda_{\mathrm{reg}} = 10^{-4}$ and Nyquist-rate direction sampling. Within the corrected directions, the element residual falls from $\sim$0~dBV to below $-80$~dBV, demonstrating effective mitigation of mutual coupling through antenna-level correction.}
    \label{fig:element-level-fullsky}
\end{figure*}

We fit the correction to the patterns at $M = \lceil 1.02 \times 2\pi/\delta^2 \rceil$ directions above the horizon, placed on an equal-area (spherical Fibonacci) lattice so that no two neighbouring directions are more than $\delta$ apart; for the perturbed Vogel station, $b_{\max} \approx 38$~m, giving $M \approx 6130$ at 122~MHz, rising to 7400 at 134~MHz. Equal-area sampling oversamples in direction cosines by a factor $1/n_z$ towards the horizon; within the constrained field of view defined below, $n_z \geq 0.96$, and $\mathbf{D}$ is zero elsewhere, so this excess does not affect the fit to an exact model (but see Section~\ref{sec:meas_noise}). For a perfectly known beam, the fit of Section~\ref{sec:element-level-cal} can then in principle reproduce the target to numerical precision within the constrained directions. In practice, the accuracy achieved also depends on how densely the sky is sampled relative to this rate and on the degree of regularisation. We therefore examine how both affect the antenna-level construction.

Figure~\ref{fig:EP_slices_sampling_and_regularisation} shows a $\mathrm{y}$-polarised cut at $\phi = 0^\circ$ and 122~MHz through the element-pattern magnitude $|\mathbf{f}_n(f,\hat{\mathbf{s}})|$ for the IEP (dashed black), the EEPs (solid grey), and their average, the average element pattern (AEP; solid black). Here and below, we call the magnitude of a far-field pattern as a function of direction and frequency, whether of an element, $|\mathbf{f}_n(f,\hat{\mathbf{s}})|$, or of a station beam, $|\mathbf{A}_l(f,\hat{\mathbf{s}})|$, its \textit{beam transfer function} (BTF). The residual between the AEP and the IEP (red) quantifies the pattern mismatch before and after correction. Although distinct from the array pattern, the AEP is proportional to it at zenith, making it a useful proxy for correction performance. Before correction (top panel), the EEPs form an envelope that reveals element-to-element spatial variation in the far-field radiation pattern due to mutual coupling and array irregularities, with residuals of order $-20$~dBV across the field of view.

\justify{The directional weighting matrix $\mathbf{D}$ is binary; we define the field of view angle $\theta_\mathrm{fov}$, which defines the non-zero weights centred on the pointing (here, zenith). In each channel, $\theta_\mathrm{fov}$ is chosen to enclose at most $N_a$ directions, the condition under which the fit reproduces the target exactly in those directions (Section~\ref{sec:element-level-cal}). Because the number of Nyquist-spaced directions within a fixed radius grows as $f^2$, $\theta_\mathrm{fov}$ falls as $1/f$, from $16.6\degr$ at 122~MHz to $15.1\degr$ at 134~MHz; a correction sharing one field of view across channels is limited by the top of the band. We computed the correction matrix of Equation~\eqref{eq:element_ls_solution} at half and full Nyquist sampling rates and applied it to the EEPs, as shown in the second and third panels, respectively. Although the half-Nyquist system remains full-rank within the correction region, it undersamples the resolvable modes, and the solution oscillates between sampled directions, a signature of spatial aliasing. At the Nyquist rate, the beam structure within the corrected region is fully resolved, and the target response is reconstructed to numerical precision, with the residual saturating at $-125$~dBV.}

In both cases, however, the unregularised solution is ill-conditioned outside the correction region, amplifying fine-scale sidelobe structure to levels upwards of 35~dBV. We therefore recompute the correction using Equation~\eqref{eq:regularised} with $\lambda_{\mathrm{reg}} = 10^{-2}$ and $10^{-4}$ (bottom two panels, respectively). Regularisation damps the poorly constrained modes associated with small singular values, which govern the response outside the correction region, and suppresses the unregularised solution's amplified sidelobes. The cost is fidelity within the correction region, which falls as $\lambda_{\mathrm{reg}}$ grows: $\lambda_{\mathrm{reg}} = 10^{-4}$ keeps the residual there below $-80$~dBV, whereas $\lambda_{\mathrm{reg}} = 10^{-2}$ gives a smoother full-sky response at the expense of a poorer fit. Regularisation therefore provides a controllable balance between sidelobe conditioning and reconstruction accuracy.

Building on \citet{o2025uncovering}, where we showed that accurate forward modelling for foreground removal requires station array patterns accurate to at least four significant figures, we require the in-field residual to lie at or below roughly $-80$~dBV and adopt $\lambda_{\mathrm{reg}} = 10^{-4}$ for the antenna-level correction. We use it throughout, except in Section~\ref{sec:requirements}, where we set $\lambda_{\mathrm{reg}} = 0$ to isolate the effect of model errors. From here on, $(\cdot)^{\mathrm{corr}}$ denotes a response after the correction has been applied.

\justify{Figure~\ref{fig:element-level-fullsky} shows the resulting full-sky responses: the upper panels show the IEP, EEP, corrected EEP, and their residuals for the central SKALA4 antenna, and the lower panels show the corresponding array patterns formed with uniform beamforming weights. The IEP varies smoothly across the sky, whereas the EEP shows pronounced mottled structure due to mutual coupling, with IEP--EEP residuals exceeding $-20$~dBV in localised regions (fourth column). After correction, the EEP recovers the IEP's smooth main-lobe morphology within the corrected region, while outside it the response tapers toward low elevations, suppressing it by about 15~dB. This taper directly follows the regularisation: where $\mathbf{D}$ imposes no constraint, the solution is governed by weakly constrained modes with small singular values, which the regularised inversion suppresses, driving the response toward zero. Within the corrected region, the residual falls below $-80$~dBV (fifth column), although the colour scale saturates near the beam centre; the remaining structure further out is dominated by the regularisation-induced taper.}

The station-level array patterns behave similarly. All display stochastic sidelobe structure with peak levels of order 20~dBV, but the EEP and corrected-EEP patterns show systematically brighter sidelobes than the IEP, particularly at low elevations. This is because element-to-element variations, whether from mutual coupling or ill-conditioning, add coherent structure that does not average down across the station. Mutual coupling contributes residuals of ${\sim}25$~dBV within the main lobe and ${\sim}15$--$20$~dBV in the sidelobes (fourth column). After correction, the element-level restoration carries through to the station beam, recovering the main-lobe morphology of the isolated case to beyond $-40$~dBV, although sidelobe residuals of $15$--$20$~dBV remain in the uncorrected directions. A local antenna-level correction thus propagates through beamforming to the full station response; Appendix~\ref{app:station_vis} shows that the same holds when the correction is applied to the beamformed voltages or the visibilities.

Evaluating Equation~\eqref{eq:regularised} requires forming the $2N_a \times 2N_a$ Gram matrix of $\mathbfcal{J}^{\mathrm{mod}}\mathbf{D}$ and its cross-product with $\mathbfcal{J}^{\mathrm{tgt}}\mathbf{D}$, each in $\mathcal{O}(M N_a^2)$ operations, and inverting the regularised Gram matrix in $\mathcal{O}(N_a^3)$; the pseudoinverse of Equation~\eqref{eq:element_ls_solution}, computed via singular value decomposition, scales in the same way. With all $M$ directions retained, and $M \propto f^2$ to keep neighbouring directions within $\delta$, the cost therefore grows quadratically with frequency. At 122~MHz, computing the correction matrix on a single 2.60~GHz core of a 38-core Intel Xeon Platinum 8368Q processor takes 2.94~s, averaged over 100 realisations. Because a binary $\mathbf{D}$ zeroes every column outside the constrained field of view, those columns contribute nothing to either equation. They can be dropped, leaving at most $N_a$ directions and a cost of $\mathcal{O}(N_a^3)$ per channel, independent of frequency. These timings exclude evaluating the radiation patterns, which we treat as a pre-computation step since its cost depends on the beam evaluation method. Once computed, $\mathbf{C}_{\mathbfcal{J}}$ folds into the beamforming weights, adding no per-sample cost and leaving downstream calibration and imaging unchanged; within the constrained field of view, those stages can model each station beam from the target pattern and array factor rather than from $N_a$ embedded patterns.

\subsubsection{Chromaticity of the Corrected Pattern}
\label{sec:spectral}

\begin{figure}
\centering
\includegraphics[width=0.97\linewidth]{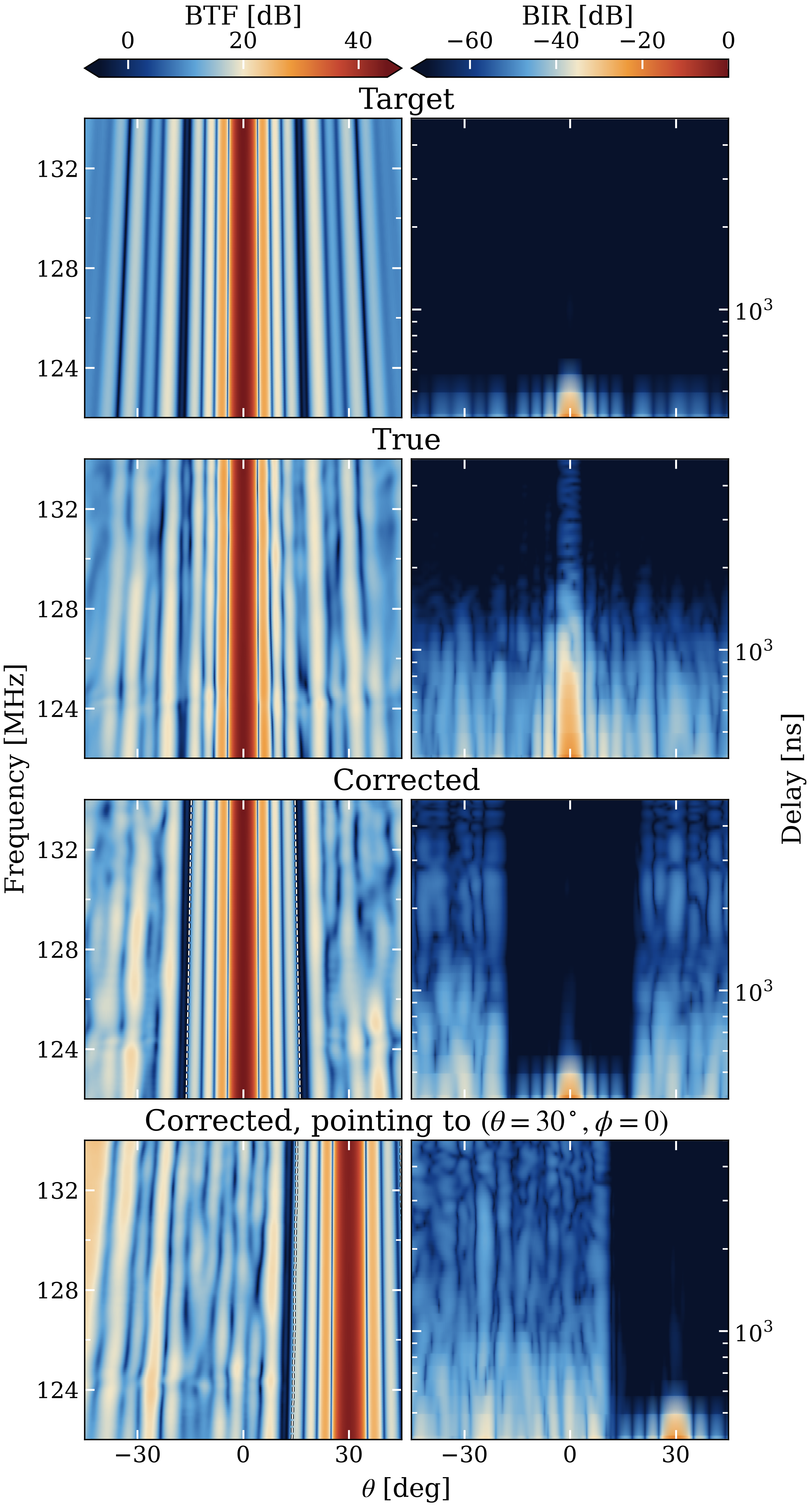}
\caption{Beam transfer function (BTF, left) and beam impulse response (BIR, right) of the X-polarised station beam of the perturbed Vogel SKA-Low station across 122--134~MHz, shown as a cut at $\phi = 0^\circ$. From top to bottom: the target beam, formed from isolated element patterns; the true beam, formed from embedded element patterns; the corrected beam, formed after antenna-level correction ($\mathbf{C}_\mathbfcal{J}$, $\lambda_{\mathrm{reg}} = 10^{-4}$), pointed at zenith; and the corrected beam pointed towards $(\theta = 30^\circ, \phi = 0^\circ)$, with the constrained field of view re-centred on the pointing. Dashed lines mark the edges of the constrained field of view. The BIR is the Fourier transform of the BTF along frequency, normalised to its peak.}
\label{fig:spectral}
\end{figure}

\justify{Power-spectrum analyses aim to separate foregrounds from the 21-cm signal based on their spectral smoothness; therefore, we judge the correction not only by how closely it matches the target in each channel, but also by how it alters the beam's spectral structure across the band. We therefore examine the station BTF as a function of frequency, along with its Fourier dual, the beam impulse response (BIR). The BIR shows directly the delays to which emission from each direction is spread: power confined to low delays keeps foreground emission at low $k_\parallel$, whereas power at high delays contaminates the EoR window. Figure~\ref{fig:spectral} shows both for the X-polarised beam of the perturbed Vogel station across 122--134~MHz, with the target array pattern formed from the IEPs, the true pattern from the EEPs, and the corrected pattern from the EEPs after the antenna-level correction $\mathbf{C}_\mathbfcal{J}$ ($\lambda_{\mathrm{reg}}=10^{-4}$).}

The target beam's chromaticity arises primarily from the array factor scaling with wavelength: its sidelobes trace straight lines across the band, and its BIR is confined to delays below a few hundred nanoseconds. Mutual coupling breaks this regularity. In the true beam, the sidelobes ripple with frequency, and the BIR extends beyond $10^3$~ns in every direction at $-40$ to $-60$~dB, while the main lobe stays within $-20$ to $-30$~dB out to about $10^3$~ns. In an interferometric observation, it is these high-delay terms that scatter foreground power into the EoR window.

\justify{Within the constrained field of view, the corrected beam recovers the target: its sidelobes again trace straight lines, and its BIR above a few hundred nanoseconds falls below $-80$~dB. Outside it, the BTF structure resembles that of the true beam; however, the correction contributes structure of its own. The correction is solved independently in each 100~kHz channel, and in the directions excluded by $\mathbf{D}$ it is dominated by the weakly constrained singular modes of $\mathbfcal{J}^{\mathrm{mod}}\mathbf{D}$. With no constraint linking neighbouring channels, these modes' orientation in the beam manifold varies from channel to channel, imprinting oscillatory structure on the unconstrained sidelobes and extending the contaminating structure to the highest delays. The correction therefore removes coupling-induced leakage only for emission within the constrained field of view; emission from outside it, such as bright sources in the far sidelobes, remains chromatic.}

\justify{Steering the beam towards $(\theta = 30^\circ, \phi = 0^\circ)$, with the constrained directions re-centred on the pointing, moves the corrected region with it, reproducing the zenith behaviour about the new pointing. Because the number of directions the correction can reproduce exactly is bounded by $N_a$ (Section~\ref{sec:element-level-cal}), the constrained field of view cannot be widened to cover every pointing at once; the correction must instead be derived for each pointing, or for each beam of a multi-beam observation.}

\subsection{Effect of Mutual-Coupling Correction on EoR Window Recovery}
\label{sec:mainbeam_sidelobe}

\begin{figure*}
    \centering
    \includegraphics[width=\linewidth]{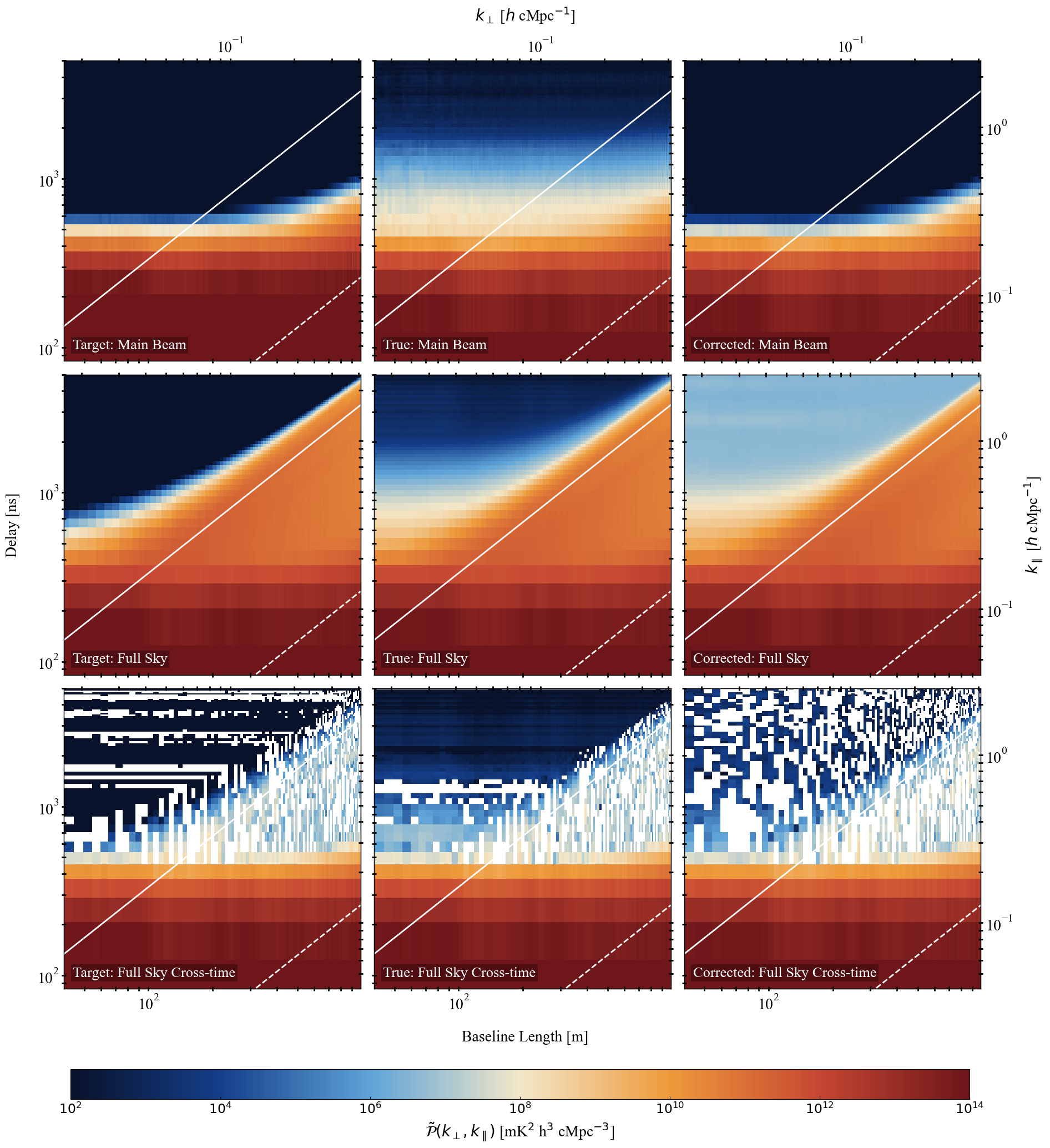}
    \caption{Two-dimensional delay power spectra for 4~hr observation tracking the EoR0 field at $(\mathrm{RA}, \mathrm{Dec.}) = (0\,\mathrm{h}, -30^{\circ})$ over 122--134~MHz, with 121 channels of 100~kHz and a sky model of GLEAM point sources and the GSM within $5\degr$ of the phase centre. Columns correspond to the target (left), true (centre), and corrected (right) station responses. In the top row, we restrict the sky to sources within $5^{\circ}$ of the phase centre to isolate the main-beam contribution; the middle and bottom rows include the full-sky source population, revealing the sidelobe contribution. The top two rows use the per-baseline estimator and the bottom row the cross-time estimator, in which white pixels denote negative estimates. Solid and dashed white lines mark the horizon and beam limits, respectively. Within the main beam, the correction suppresses coupling-induced leakage from the true response and recovers the target's EoR window. For the full sky, the unconstrained sidelobes of the corrected response raise the power in the EoR window above that of the uncorrected response in the per-baseline estimate; in the cross-time estimate, this excess decorrelates between correction cycles and scatters about zero, whereas the leakage of the true response remains coherent.}
    \label{fig:Cal_mainbeam_vs_fullsky}
\end{figure*}

\begin{figure*}
    \centering
    \includegraphics[width=.9\linewidth]{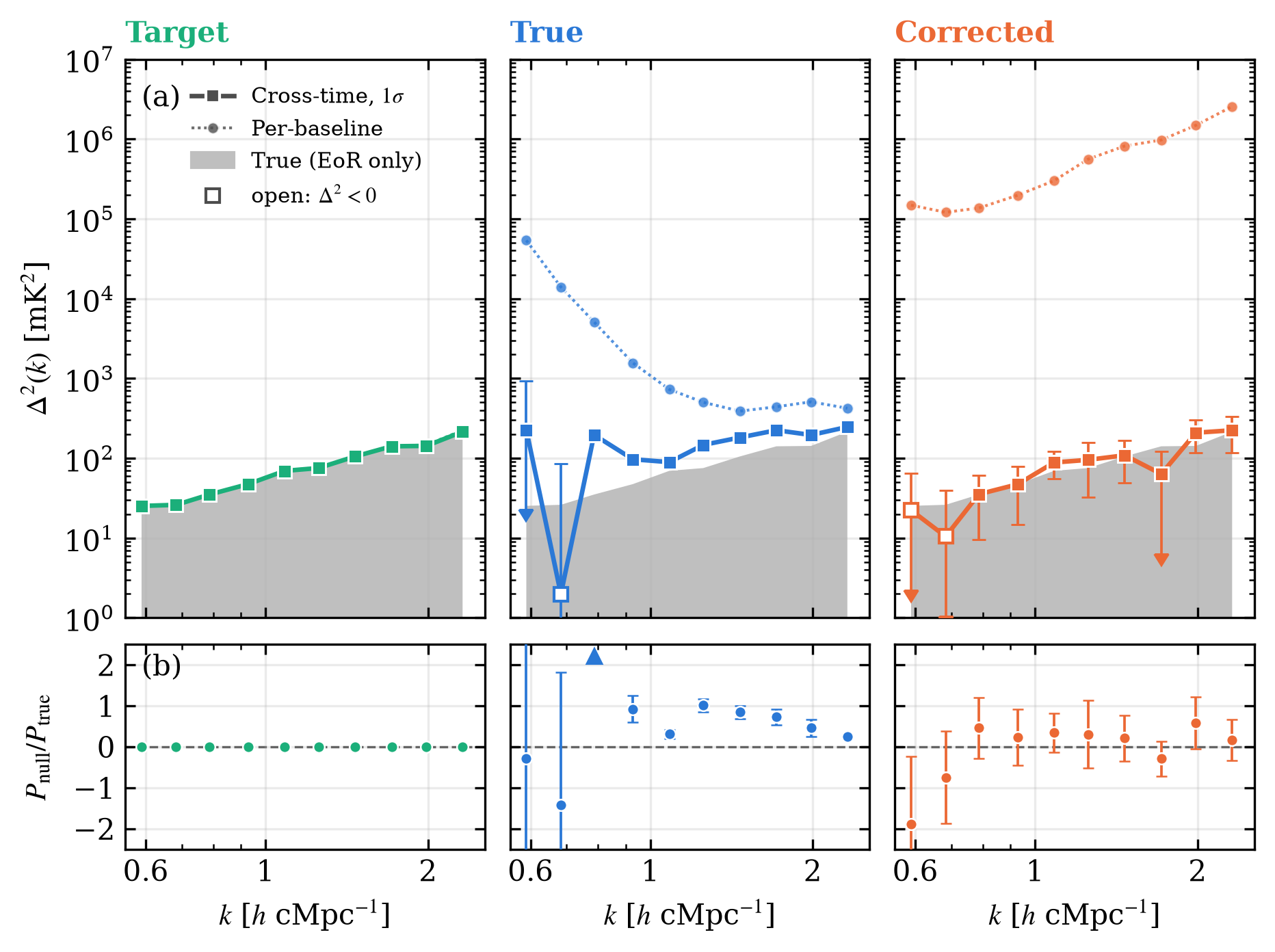}
    \caption{Spherically averaged dimensionless power spectra for simulated observations of the EoR0 field over 122--134~MHz, with a sky comprising GLEAM point sources, GSM diffuse emission and the 21-cm signal. Columns correspond to the target (left), true (centre), and corrected (right) station responses. (a)~Cross-time (solid) and per-baseline (dotted) estimates of $\Delta^2(k)$, compared with the input 21-cm power spectrum (grey). Cross-time error bars are $1\sigma$, estimated from a delete-one-group jackknife over 12 groups of correction cycles, and arrows mark intervals that extend below zero; per-baseline estimates are shown without error bars. Open markers denote bins in which $\Delta^2 < 0$, plotted as $|\Delta^2|$. (b)~Null test, in which the cross-time estimator is applied to a foreground-only sky, expressed as a fraction of the input 21-cm power, $\Delta^2_{\mathrm{null}}/\Delta^2_{\mathrm{true}}$, with the $1\sigma$ jackknife errors of the foreground+EoR estimate; triangles mark values beyond the plotted range. All errors are statistical and exclude systematics coherent between correction cycles, which the null test probes. The corresponding recovery statistics are given in Table~\ref{tab:pk1d}.}
    \label{fig:pk1d}
\end{figure*}

\justify{Section~\ref{sec:spectral} showed that, within the constrained field of view, the correction restores the spectral smoothness of the station beam. We now ask how this carries through to the 21-cm power spectrum, and whether the cross-time estimator can remove coupling-induced leakage without it. For each of the target, true and corrected station responses, we simulate the observation of Section~\ref{sec:sim_setup} for two foreground skies, each combining the GSM within $5\degr$ of the phase centre with GLEAM sources either within the same radius, isolating the main lobe where the correction is constrained, or above the horizon, adding the sidelobes where it is not. Figure~\ref{fig:Cal_mainbeam_vs_fullsky} shows the resulting two-dimensional power spectra, and Figure~\ref{fig:pk1d} shows the spherically averaged power spectra with the 21-cm signal added, together with a foreground-only null test. Power is quoted in $\mathrm{mK}^2\,h^{3}\,\mathrm{cMpc}^{-3}$ throughout.}

\justify{Within the main lobe (Figure~\ref{fig:Cal_mainbeam_vs_fullsky}, top row), the target response confines the foreground power to the lowest delays and the wedge, with the power falling below $\sim10^{2}$ beyond $\sim600$~ns on the shortest baselines. Mutual coupling in the true response spreads this power to higher delays on every baseline, with power above $\sim10^{4}$ extending to $\sim2~\mu$s. Its horizontal banding, independent of baseline length, shows that the leakage originates in the station beam's spectral structure (Figure~\ref{fig:spectral}), rather than in the baseline-dependent chromaticity that forms the wedge. This leakage is weaker than in the 120--150~MHz band of \citet{o2025uncovering}, as fewer of the sharp coupling-induced resonances fall within the narrower 12~MHz band, but it remains far above the 21-cm signal. The corrected response closely reproduces the target, with a sharply bounded wedge and a clean EoR window. Within its constrained field of view, the correction removes the coupling-induced spectral structure.}

\justify{Including the full sky (middle row) changes the picture. The target wedge now fills to the horizon, with supra-horizon emission extending to $\sim1~\mu$s on the shortest baselines, and the true response again leaks power into the EoR window, here at $\sim10^{4}$--$10^{6}$ and decreasing with delay. The corrected response, however, raises the EoR window to $\sim10^{6}$--$10^{7}$, nearly uniformly in delay and above the uncorrected leakage. This excess originates in the sidelobes. Directions outside the constrained field of view are excluded by $\mathbf{D}$, so the correction there is governed by the weakly constrained singular modes of $\mathbfcal{J}^{\mathrm{mod}}\mathbf{D}$. The independent per-channel correction results in these modes varying erratically from channel to channel across the band (Section~\ref{sec:spectral}), imprinting spectrally incoherent structure on sidelobe emission that spreads evenly across delay.}

\justify{The cross-time estimate (bottom row) separates the two. The coupling-induced leakage of the true response is a property of the instrument that changes only slowly as the field is tracked. Hence, a portion of it remains coherent between blocks and survives as structured positive power of $\sim10^{4}$--$10^{6}$ between $\sim0.6$ and $1.2~\mu$s on the shortest baselines. The sidelobe excess of the corrected response, by contrast, is not reproduced between correction cycles. As the field is tracked, the constrained field of view moves across the station response, so the unconstrained sidelobe modes, and the excess they produce, change from cycle to cycle. The excess therefore decorrelates and scatters about zero, with negative estimates (white) interspersed throughout the EoR window.}

\justify{Figure~\ref{fig:pk1d} and Table~\ref{tab:pk1d} quantify the consequences for recovering the 21-cm power spectrum. We summarise each cross-time estimate by its recovered amplitude $A$, the least-squares scaling of the input 21-cm power spectrum that best fits it, $\Delta^2(k) = A\,\Delta^2_{\mathrm{true}}(k)$, over $k > 0.85\,h\,\mathrm{cMpc}^{-1}$; this excludes the lowest-$k$ bins, which lie closest to the wedge and are poorly constrained due to limited samples. We compare the estimate with the input 21-cm power, and the foreground-only estimate with zero as a null test, through $\chi^2$ over all 10 $k$ bins with the $1\sigma$ jackknife errors, and report its probability to exceed (PTE), the chance of a $\chi^2$ at least this large if the estimate were consistent with the input; an unbiased recovery therefore gives $A \approx 1$ and a PTE that is not small. The target response recovers the input exactly ($A = 1.00 \pm 0.01$). For the true response, the per-baseline estimate is foreground-dominated, exceeding the input by three orders of magnitude at $k \approx 0.6\,h\,\mathrm{cMpc}^{-1}$ and by a factor of two to three at the highest $k$. The cross-time estimator removes most of this power. Still, the coherent leakage survives: the recovered amplitude is biased high, $A = 1.31 \pm 0.08$, the estimate is inconsistent with the input ($\chi^2 = 117.9$, PTE $<10^{-10}$), and the null test detects residual foreground power at 25--100 per cent of the 21-cm signal above $k \approx 0.9\,h\,\mathrm{cMpc}^{-1}$ ($\chi^2 = 146.2$, PTE $<10^{-10}$). For the corrected response, the per-baseline estimate is worse still, $10^{3}$--$10^{4}$ times the input and rising with $k$, consistent with power spread across all delays. Yet its cross-time estimate is consistent with the input ($A = 1.00 \pm 0.18$; $\chi^2 = 5.5$, PTE $= 0.85$), and its null test with zero ($\chi^2 = 4.5$, PTE $= 0.92$).}

\begin{table}
\centering
\caption{Recovery of the input 21-cm power spectrum by the cross-time estimator for the target, true, and corrected responses of Figure~\ref{fig:pk1d}. The recovered fraction $A$ is the least-squares amplitude of $\Delta^2(k) = A\,\Delta^2_{\mathrm{true}}(k)$ over $k > 0.85\,h\,\mathrm{cMpc}^{-1}$, with $1\sigma$ jackknife errors. The $\chi^2$ tests compare the cross-time estimate with the 21-cm power of the input sky, which is $\Delta^2_{\mathrm{true}}$ for the foreground+EoR sky (input) and zero for the foreground-only sky (null), using the jackknife errors of the foreground+EoR estimate in both cases. We compute each $\chi^2$ over all 10 bins ($\nu = 10$) assuming diagonal errors, and report its PTE. The tests are not quoted for the target response, whose estimate departs from the input only at the level of numerical precision, far below the jackknife errors.}
\begin{tabularx}{\columnwidth}{@{}l*{3}{>{\centering\arraybackslash}X}@{}}
\toprule
 & \textbf{Target} & \textbf{True} & \textbf{Corrected} \\
\midrule
\midrule
 $A$ ($k > 0.85$)   & $1.00 \pm 0.01$ & $1.31 \pm 0.08$ & $1.00 \pm 0.18$ \\
\midrule
 $\chi^2$, input    & --              & $117.9$         & $5.5$  \\
 PTE, input         & --              & $<10^{-10}$     & $0.85$ \\
\midrule
 $\chi^2$, null     & --              & $146.2$         & $4.5$  \\
 PTE, null          & --              & $<10^{-10}$     & $0.92$ \\
\bottomrule
\end{tabularx}
\label{tab:pk1d}
\end{table}

\justify{The correction thus converts a coherent bias into variance. The decorrelated sidelobe power enters the cross-time estimate as additional scatter, more than doubling the uncertainty on $A$ relative to the true response. Still, unlike the bias of the uncorrected instrument, this scatter is statistical and should diminish as more independent correction cycles are combined. These results assume an exact EEP model, whose required accuracy Section~\ref{sec:requirements} quantifies, and a sky of GLEAM point sources across the visible hemisphere with diffuse emission restricted to within $5\degr$ of the phase centre. Diffuse emission in the sidelobes, where the correction is unconstrained, is likely to increase the variance; we leave a full-sky diffuse test, pixelised at the resolution of the core baselines, to future work. A reconstructed power-spectrum approach \citep{morales_diversity}, which forms images before the line-of-sight transform and so allows sidelobe emission to be apodised, could reduce this penalty. In contrast, Gaussian Process Regression \citep{liu_gpr, mertens_gpr} could remove the residual smooth foregrounds that limit the lowest-$k$ bins, although we have not tested either here.}

\justify{Our tests are also limited to a single station configuration, although nothing in the correction is specific to it. $\mathbf{C}_\mathbfcal{J}$ is fitted directly to the modelled patterns, without a coupling matrix or the single-mode assumption of \citet{steyskal1990mutual}, so it applies regardless of the physical origin of the pattern distortion. We therefore expect the same behaviour for other elements, layouts and frequency bands whose EEPs are accurately modelled. However, how closely the target is recovered will depend on the conditioning of $\mathbfcal{J}^{\mathrm{mod}}\mathbf{D}$ for each design.}

\subsection{Robustness to Beam-Model Errors}
\label{sec:robustness}

\begin{figure}
\centering
\includegraphics[width=0.88\linewidth]{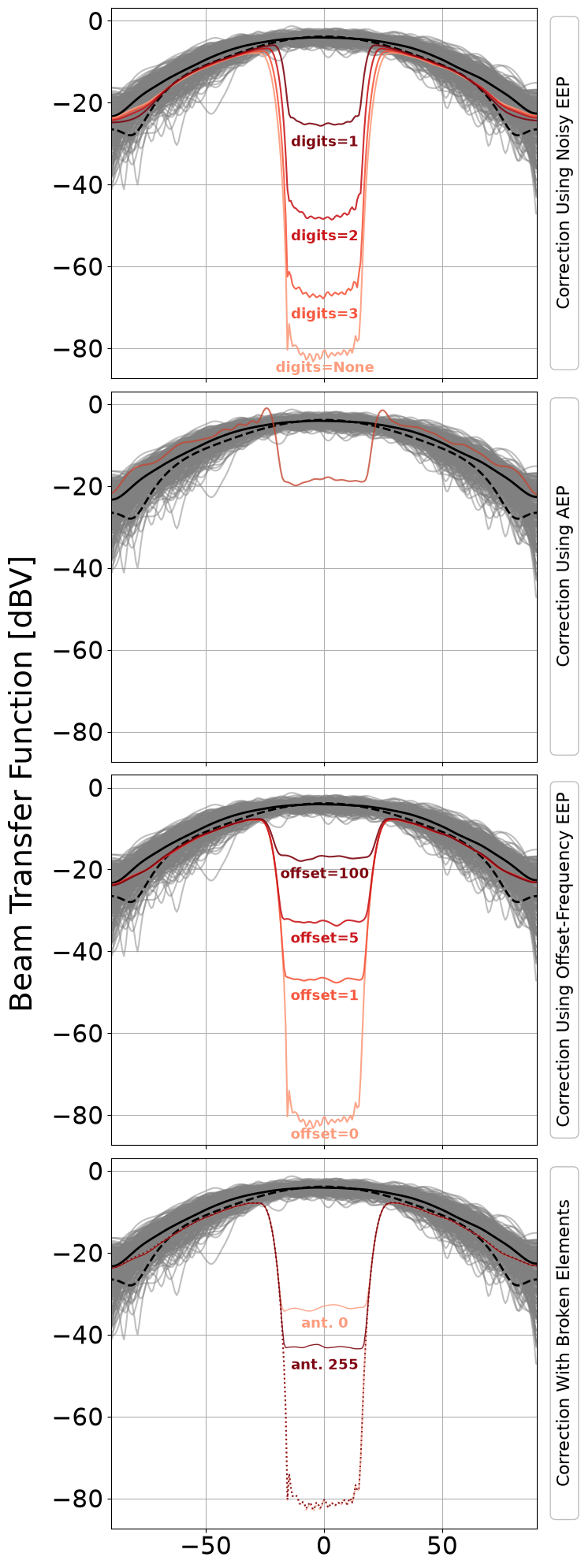}
\caption{Sensitivity of antenna-level mutual coupling compensation to errors in the EEP model, for the perturbed Vogel SKA-Low station at 122~MHz. Each panel shows a $\phi = 0^\circ$ cut through the Y-polarisation of all EEPs (grey), their average (solid black), the isolated element pattern (dashed black), and the residual between the corrected and target responses (coloured), for the model error labelled at right; `None' and `offset=0' denote the unperturbed model. Channel offsets are in units of 100~kHz. Antennas 0 and 255 lie at the centre and edge of the station, and dotted lines show the residual once the failed element is accounted for in the model.}
\label{fig:noise}
\end{figure}

\justify{In practice, however, the modelled far-field pattern used to construct the correction is never exact. Simulated patterns carry numerical errors, are computed at finite spectral resolution, and assume that the as-built instrument matches its design \citep{cumner2024effects}; patterns inferred from calibration solutions inherit errors from model mis-specification \citep{brackenhoff2025robust}; measured patterns carry observational uncertainty; and, during operation, degrading LNAs or cables can cause elements to fail \citep{joseph2020calibration} or imprint per-element chromatic gains through reflections and length tolerances \citep{ohara_under}. We test the correction against four classes of model error, shown in the four panels of Figure~\ref{fig:noise}: Gaussian perturbations of controlled precision, substitution of the Average Element Pattern (AEP), EEPs taken from an offset frequency channel, and a failed element. In each case, the correction $\mathbf{C}_\mathbfcal{J}$ is solved via Equation~\eqref{eq:regularised} from a model $\mathbfcal{J}^{\mathrm{mod}}$ that departs from the true response, and is then applied to the true response $\mathbfcal{J}^{\mathrm{true}}$, reproducing the operational scenario in which the instrument is modelled imperfectly. The Nyquist-sampled, $\lambda_\mathrm{reg}=10^{-4}$ solution shown in the fifth panel of Figure~\ref{fig:EP_slices_sampling_and_regularisation} serves as the control, with residuals below $-80$~dBV within the constrained field of view.}

\subsubsection{Gaussian Model Noise}
\label{sec:meas_noise}

While the error characteristics of a real beam model depend on its origin, a simpler question is what numerical precision the EEPs require for the correction to remain effective. We add independent complex Gaussian noise to each modelled EEP,
\begin{equation}
    \mathbf{f}^{\mathrm{mod}}_{n}(f, \hat{\mathbf{s}}) = \mathbf{f}^{\mathrm{true}}_{n}(f, \hat{\mathbf{s}}) + \mathbfcal{N}_n(f, \hat{\mathbf{s}}),
\end{equation}
where each entry of $\mathbfcal{N}_n$ is drawn from $\mathcal{CN}(0,\sigma^2)$, the complex normal distribution with zero mean and variance $\sigma^2$, independently for each antenna $n$, frequency $f$, and direction $\hat{\mathbf{s}}$. Rather than modelling a specific physical error source, $\sigma^2$ is chosen to degrade the model to a prescribed number of significant figures, providing a controlled proxy for overall model accuracy.

\justify{The residual floor rises monotonically as precision falls (Figure~\ref{fig:noise}, top panel), from below $-80$~dBV for the unperturbed model to approximately $-65$, $-47$, and $-25$~dBV at three, two, and one significant figures, respectively. Noise perturbs the singular spectrum of the weighted model $\mathbfcal{J}^{\mathrm{mod}}\mathbf{D}$, and hence the regularised pseudoinverse; as the perturbation shrinks, the correction converges to the idealised solution. Outside the constrained field of view, the response depends mainly on the weakly constrained singular modes of $\mathbfcal{J}^{\mathrm{mod}}\mathbf{D}$. Noise lowers the effective signal-to-noise ratio of these modes, so the regularisation suppresses them more strongly, biasing the solution towards the minimum-norm estimate and deepening the taper towards zero already present in the noise-free correction (Figure~\ref{fig:element-level-fullsky}, top row, third column).}

\subsubsection{AEP Substitution}
\label{sec:aep}

\justify{Calibration procedures commonly assume that every element shares a single pattern, as noted in the Introduction. We test this approximation by assigning the AEP to every element of the model. The residual within the constrained field of view rises to about $-19$~dBV (Figure~\ref{fig:noise}, second panel), worse than a model accurate to only one significant figure, and peaks near $-1$~dBV at the edge of the constrained region, exceeding the average pattern itself. This occurs because AEP discards the element-to-element variation introduced by mutual coupling; the correction has no information with which to remove it.}

\subsubsection{Spectral Channel Offsets}
\label{sec:offsets}

\justify{Because full-wave simulations are computationally expensive, EEPs are often computed on a coarser frequency grid than the observation and applied at the nearest available frequency. We test this by constructing the model from EEPs offset by 1, 5, and 100 channels (100~kHz each) from the observing frequency. A single-channel offset raises the residual floor to about $-47$~dBV, equivalent to a two-significant-figure model, while offsets of 5 and 100 channels raise it to about $-33$ and $-18$~dBV (Figure~\ref{fig:noise}, third panel), the latter comparable to the AEP case. This sensitivity reflects the sub-MHz spectral structure that mutual coupling imprints on the EEPs: a correction derived in one channel does not transfer to its neighbours, so the EEPs must be modelled at the spectral resolution at which the correction is applied.}

\subsubsection{Broken Elements}
\label{sec:failed}

\justify{Finally, we consider a single element that fails in the instrument but remains in the model used to construct the correction. We assume the failure occurs in the RF-over-fibre link downstream of the LNA, so the antenna and its termination are unaffected: the remaining elements' EEPs are unchanged, and only the failed element's signal is lost. A failure at antenna 0, at the centre of the station, raises the residual floor to about $-33$~dBV, and a failure at antenna 255, on its edge, to about $-43$~dBV (Figure~\ref{fig:noise}, bottom panel). This is consistent with the central element coupling to more neighbours, so its signal corrects more elements. Once the failure is identified by quality-assurance metrics and the element is removed from the model, the re-derived correction returns the residual to the control level (dotted curves), so the method tolerates element failures provided they are flagged.}

\justify{Together, these tests place the error sources on a common scale. Substituting the AEP, or using EEPs 100 channels from the observing frequency, limits the correction to roughly the level of a one-significant-figure model, while even a single-channel offset is equivalent to two significant figures. The correction is therefore only as good as the model's ability to capture element-to-element and channel-to-channel variation; approximations that average over either forfeit most of its benefit.}

\subsection{Model Requirements for 21-cm Detection}
\label{sec:requirements}

Section~\ref{sec:robustness} illustrates the correction accuracy under model misspecification for a given channel; however, this alone does not determine the impact on the 21-cm power spectrum, which depends instead on the residual's chromaticity across the band. To quantify this, we now propagate the model errors through to the power spectrum to assess the impact of their spectral structure. To isolate the impact of these errors on the residual foreground contamination, we impose $\lambda_{\mathrm{reg}}=0$ and restrict the constrained directions such that the system is full rank, thereby making the correction a direct inversion of the model response; given an exact model, it reproduces the target exactly as the minimum-norm solution, whose residuals reach the numerical floor. The simulated sky model is composed of the GLEAM catalogue and restricted to the constrained field of view, so directions where the correction is unconstrained contribute no foreground power. While such a direct inversion is not the most robust choice when the model itself is in error, we aim to derive benchmarks for the beam knowledge that a linear correction requires, against which other mitigation strategies can be measured.

For this, we opt for two Gaussian error models: the first being the model of Section~\ref{sec:meas_noise}, whose errors are independent between channels, and another whose errors are correlated in frequency, drawn as a Gaussian random process with correlation $\rho(\Delta f) = \exp[-\Delta f^2/(2\Delta f_\mathrm{c}^2)]$ between channels $\Delta f$ apart, with $\Delta f_\mathrm{c} = 1$~MHz. In both cases, we draw the errors on a per-element, per-polarisation, and per-direction basis, with a variance constant between channels so that we can map the noise directly to pattern accuracy in significant figures. For each level and error model, we simulate visibilities with \texttt{OSKAR} across the band, apply the correction, and compute the signal-to-foreground ratio (S/F) of the 21-cm signal within the EoR window for the shortest baseline bin.

\begin{figure}
\centering
\includegraphics[width=\linewidth]{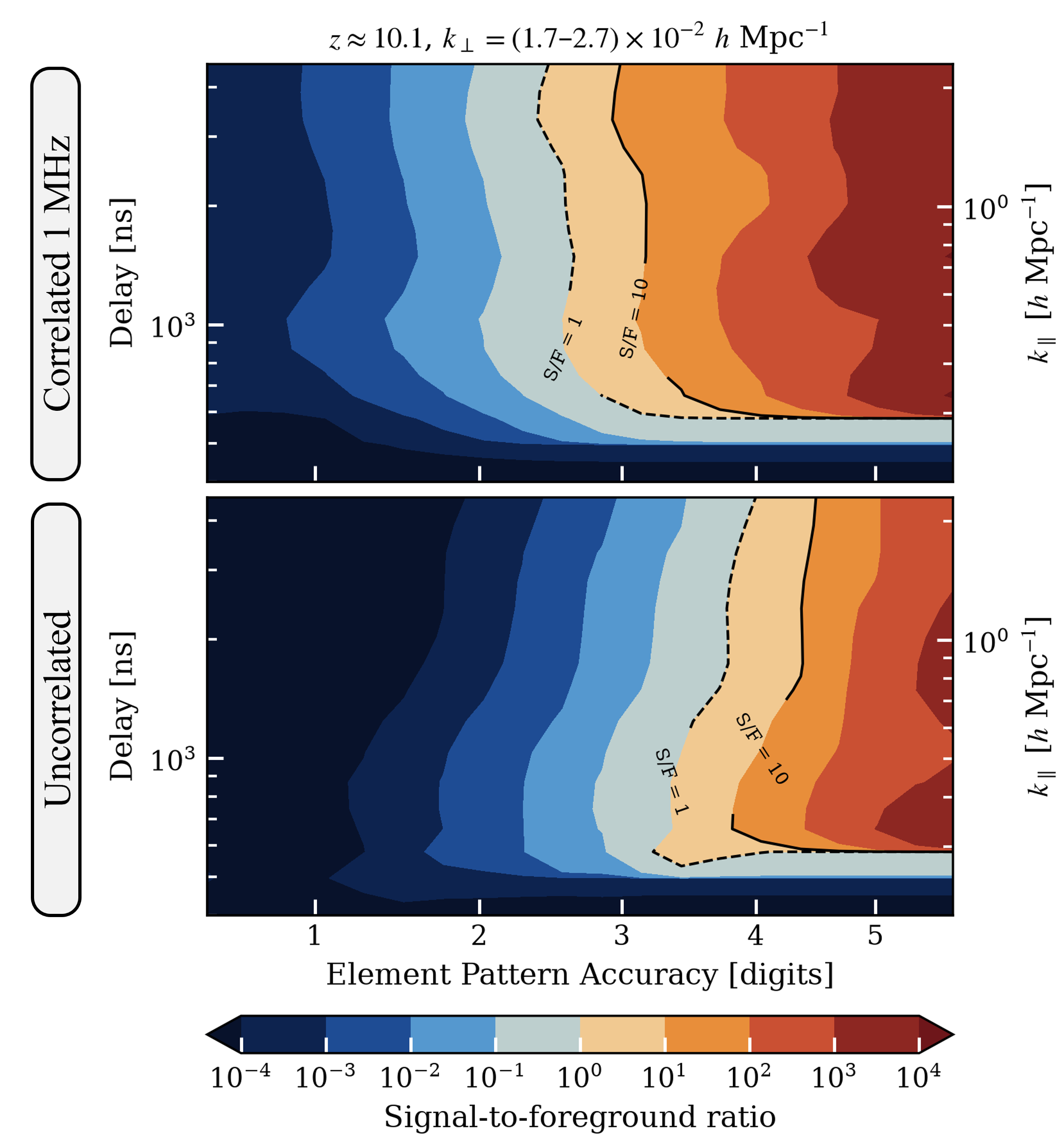}
\caption{Signal-to-foreground ratio within the EoR window for the shortest baseline bin, $k_\perp = (1.7$--$2.7)\times10^{-2}\,h\,\mathrm{cMpc}^{-1}$ at $z \approx 10.1$, as a function of EEP model accuracy and delay, after antenna-level correction with $\lambda_{\mathrm{reg}} = 0$. Visibilities are simulated over 122--134~MHz for a GLEAM sky model restricted to the constrained field of view, with the EEPs degraded to between 0.5 and 5.5 significant figures by complex Gaussian errors that are correlated across 1~MHz (top) or independent between channels (bottom). Dashed and solid contours mark S/F~$=1$ and S/F~$=10$, respectively.}
\label{fig:beam_error}
\end{figure}
 
\justify{At the lowest delays of Figure~\ref{fig:beam_error}, below about 600~ns, the S/F remains below unity regardless of the accuracy for both error models: given the observational parameters, this region corresponds to that of the intrinsic foregrounds, which are inherently degenerate with the 21-cm signal, which no correction to the instrument response can remove. Above this region, the S/F rises monotonically with model accuracy.}

For uncorrelated errors (bottom panel), the S/F first reaches unity at about 3.25 significant figures and 10 at about 3.75, just above the foreground-dominated region. The required accuracy is lowest there and increases towards higher delays, to about 3.9 and 4.5 figures, respectively, at 4~$\mu$s; this behaviour occurs because the errors are drawn independently per channel and the resulting foreground leakage is flat and uniform across delay, whereas the 21-cm power declines with increasing $k_\parallel$; at fixed accuracy, S/F therefore falls with delay.

\justify{When the errors are correlated in frequency, the accuracy requirements are relaxed (top panel): over most of the EoR window, S/F reaches unity at about 2.5 significant figures and 10 at about 3, some 1.5 figures below the independent-error requirement at 4~$\mu$s. However, for a Gaussian correlation width of $\Delta f_\mathrm{c} = 1$~MHz, one would expect the delay spectrum of the injected errors to fall as $\exp(-2\pi^2\Delta f_\mathrm{c}^2\tau^2)$, falling to about $-85$~dB at 1~$\mu$s. Were the leakage set by the errors alone, the EoR window would be almost free of it above 1~$\mu$s, and the accuracy requirement would all but vanish there. The injected errors alone therefore cannot account for the leakage that remains, which we attribute to the correction itself: each channel's errors pass through that channel's inverse, which carries the sub-MHz structure of the coupling, so the change they make to the corrected EEPs is less correlated across channels than the errors themselves, and the resulting leakage follows the spectral structure of the coupled response, which falls with delay (Figure~\ref{fig:spectral}). Spectrally smooth model errors thus relax the requirement, but the correction re-imprints the coupling's own spectral structure on them. Since the spectral structure of the errors in a real beam model, whether simulated, measured or derived from calibration, cannot be assumed, we take the more demanding independent-error thresholds as a worst-case scenario.}

Even under these favourable conditions, the approximations commonly used in the field fall short: assigning the AEP to every element (Section~\ref{sec:aep}), or reusing EEPs from a channel only 100~kHz away, as when a single model is applied across a block of channels (Section~\ref{sec:offsets}), corresponds to at most one and about two significant figures, respectively. This requirement is comparable to that of forward modelling and subtraction, which require EEPs accurate to three to four significant figures \citep{o2025uncovering}; subtraction, however, also requires sky models of similar accuracy, which the field does not yet possess \citep{mkay_sky_model_correction}, whereas the correction depends on the beam model alone. In practice, a decisive detection would need to meet these requirements or demonstrate a method whose performance is invariant to beam-model misspecification.

\section{Conclusions}
\label{sec:conclusions}

\justify{Mutual coupling imprints sub-MHz spectral structure on the station beams of dense aperture arrays, scattering foreground power into the EoR window. We presented direct primary beam correction, a regularised, direction-weighted linear operator that transforms an aperture array's far-field response toward an arbitrary target response. We applied it to full-wave simulations of a Perturbed Vogel SKA-Low station between 122 and 134~MHz, using the isolated element pattern as the target. This yielded the following key findings:}
\begin{enumerate}
    \item \justify{\textit{Mutual coupling can be removed accurately and at low cost, complementing direction-dependent calibration.} With the model sampled at the Nyquist rate, the correction reproduces the target far-field pattern exactly for as many directions as there are elements, which limits the constrained field of view (a radius of $16.6\degr$ at 122~MHz, narrowing to $15.1\degr$ at 134~MHz) and requires the correction to follow the beam when it is steered. In this region, regularisation sets the accuracy, trading in-field fidelity for sidelobe conditioning (for $\lambda_{\mathrm{reg}} = 10^{-4}$, the residuals stay below $-80$~dBV at 122~MHz). The correction is a single linear operator, cheap to compute (here about 2.9~s per channel at 122~MHz for the full-sky pixel set) and applied as weights during beamforming; retaining only the constrained directions reduces the cost to $\mathcal{O}(N_a^3)$ per channel, independent of frequency. Ionospheric and other time-variable effects remain the task of downstream calibration, which the correction supports by restoring the smooth, common beam it assumes.}

    \item \justify{\textit{The correction restores spectral smoothness within the constrained region.} Mutual coupling extends the beam impulse response beyond $10^{3}$~ns at $-40$ to $-60$~dB, scattering foreground power into the EoR window, even in this band, where fewer sharp coupling-induced resonances fall within the 12~MHz range than in wider bands. Within the constrained region, the corrected beam's impulse response falls below $-80$~dB after a few hundred nanoseconds, recovering that of the target. Outside it, the sidelobes are effectively uncorrected, and channel-to-channel variation in the weakly constrained modes adds spectral structure of its own, extending to the highest delays.}
    
    \item \justify{\textit{A coherent bias becomes statistical scatter.} The cross-time estimator suppresses most of the coupling leakage, but without correction a coherent residual survives, biasing the recovered 21-cm amplitude to $A = 1.31 \pm 0.08$ (PTE~$<10^{-10}$), a bias that cannot be integrated away. With correction, the channel-to-channel variation in the weakly constrained modes decorrelates between correction cycles, giving an unbiased $A = 1.00 \pm 0.18$ (PTE~$=0.85$) and a passing null test. The remaining scatter is statistical and falls as more cycles are combined.}

    \item \justify{\textit{The beam model sets the limit.} We tested the correction against four classes of model error: Gaussian perturbations of controlled precision, substitution of the average element pattern (AEP), EEPs taken from an offset frequency channel, and a failed element. Degrading the EEPs to three, two and one significant figures raises the residual floor from below $-80$~dBV to about $-65$, $-47$ and $-25$~dBV. Substituting the AEP raises it to about $-19$~dBV, and a one-channel (100~kHz) offset to about $-47$~dBV, equivalent to two significant figures. A single failed element is tolerated once it is flagged and removed from the model.}

    \item \justify{\textit{Detection requires at least 3.25 significant figures on the embedded element patterns.} Propagating Gaussian model errors in the EEPs to the power spectrum, the signal-to-foreground ratio in the EoR window reaches unity at between 3.25 and 3.9 significant figures of EEP accuracy, depending on delay, and errors correlated in frequency relax this to about 2.5. These are necessary rather than sufficient conditions. The requirement is comparable to that of forward modelling and subtraction \citep{o2025uncovering}, but the correction depends on the beam model alone, and the approximations in common use reach at most two significant figures of EEP accuracy.}
\end{enumerate}

This work shows that, given an accurate beam model, the correction removes the coherent bias mutual coupling introduces into the 21-cm power spectrum, and that beam characterisation, not the correction, limits coupling mitigation. Future work should test the cross-time estimator with imperfect beam models, thermal noise and calibration errors; extend the foreground model to the full diffuse sky; explore deliberately perturbing the correction between time steps to decorrelate the sidelobe residuals; combine it with a reconstructed power-spectrum approach; and test it on a deployed station, anchored to measured beam models.

\section*{Acknowledgements}

This research was supported by the Science and Technology Facilities Council (STFC). O.S.D. O'Hara, J. Cumner and D. Anstey were supported by grant number ST/X00239X/1, Q. Gueuning by grant number ST/Y000447/1, E. de Lera Acedo by grant number ST/V004425/1, F. Dulwich by grant number ST/W00206X/1 and A. Mhaske by grant number G127873. The work of Oskar Zetterstrom has been funded by the Swedish Research Council (project number 2024-06677).

\section*{Data Availability}
The data and software underlying this article will be shared on reasonable request to the corresponding authors.



\bibliographystyle{mnras}
\bibliography{bibliography} 

@article{van2013lofar,
  title={LOFAR: The low-frequency array},
  author={van Haarlem, Michael P and Wise, Michael W and Gunst, AW and Heald, George and McKean, John P and Hessels, Jason WT and de Bruyn, A Ger and Nijboer, Ronald and Swinbank, John and Fallows, Richard and others},
  journal={Astronomy \& astrophysics},
  volume={556},
  pages={A2},
  year={2013},
  publisher={EDP Sciences},
  doi={10.1051/0004-6361/201220873}
}

@article{deboer2017hydrogen,
  title={Hydrogen epoch of reionization array (HERA)},
  author={DeBoer, David R and Parsons, Aaron R and Aguirre, James E and Alexander, Paul and Ali, Zaki S and Beardsley, Adam P and Bernardi, Gianni and Bowman, Judd D and Bradley, Richard F and Carilli, Chris L and others},
  journal={Publications of the Astronomical Society of the Pacific},
  volume={129},
  number={974},
  pages={045001},
  year={2017},
  publisher={IOP Publishing},
  doi={10.1088/1538-3873/129/974/045001}
}

@article{craeye2011review,
  title={A review on array mutual coupling analysis},
  author={Craeye, Christophe and Gonz{\'a}lez-Ovejero, David},
  journal={Radio Science},
  volume={46},
  number={02},
  pages={1--25},
  year={2011},
  publisher={AGU},
  doi={10.1029/2010RS004518}
}

@article{bui2018direct,
  title={Direct deterministic nulling techniques for large random arrays including mutual coupling},
  author={Bui-Van, Ha and Hamaide, Valentin and Craeye, Christophe and Glineur, Fran{\c{c}}ois and de Lera Acedo, Eloy},
  journal={IEEE Transactions on Antennas and Propagation},
  volume={66},
  number={11},
  pages={5869--5878},
  year={2018},
  publisher={IEEE},
  doi={10.1109/TAP.2018.2869068}
}

@article{huang2013mutual,
  title={Mutual coupling calibration for microstrip antenna arrays via element pattern reconstruction method},
  author={Huang, Qiulin and Zhou, Hongxing and Bao, Jianhui and Shi, Xiaowei},
  journal={IEEE Antennas and Wireless Propagation Letters},
  volume={13},
  pages={51--54},
  year={2013},
  publisher={IEEE},
  doi={10.1109/LAWP.2013.2296073}
}

@inproceedings{wijnholds2020embedded,
  title={Embedded element patterns in hierarchical calibration of large distributed arrays},
  author={Wijnholds, Stefan J},
  booktitle={2020 XXXIIIrd General Assembly and Scientific Symposium of the International Union of Radio Science},
  pages={1--4},
  year={2020},
  organization={IEEE},
  doi={10.23919/URSIGASS49373.2020.9232295}
}

@inproceedings{wijnholds2019using,
  title={Using embedded element patterns to improve aperture array calibration},
  author={Wijnholds, Stefan J and Arts, Michel and Bolli, Pietro and Di Ninni, Paola and Virone, Giuseppe},
  booktitle={2019 international conference on electromagnetics in advanced applications (ICEAA)},
  pages={0437--0442},
  year={2019},
  organization={IEEE},
  doi={10.1109/ICEAA.2019.8878949}
}

@article{o2025uncovering,
  title={Uncovering the effects of array mutual coupling in 21-cm experiments with the SKA-Low radio telescope},
  author={O’Hara, Oscar Sage David and Gueuning, Quentin and de Lera Acedo, Eloy and Dulwich, Fred and Cumner, John and Anstey, Dominic and Brown, Anthony and Fialkov, Anastasia and Dhandha, Jiten and Faulkner, Andrew and others},
  journal={Monthly Notices of the Royal Astronomical Society},
  volume={538},
  number={1},
  pages={31--48},
  year={2025},
  publisher={Oxford University Press},
  doi={10.1093/mnras/staf264}
}

@article{smirnov2011revisitingI,
  title={Revisiting the radio interferometer measurement equation-I. A full-sky Jones formalism},
  author={Smirnov, Oleg M},
  journal={Astronomy \& Astrophysics},
  volume={527},
  pages={A106},
  year={2011},
  publisher={EDP Sciences},
  doi={10.1051/0004-6361/201016082}
}

@article{smirnov2011revisitingII,
  title={Revisiting the radio interferometer measurement equation-II. Calibration and direction-dependent effects},
  author={Smirnov, Oleg M},
  journal={Astronomy \& Astrophysics},
  volume={527},
  pages={A107},
  year={2011},
  publisher={EDP Sciences},
  doi={10.1051/0004-6361/201116434}
}

@article{bhatnagar2008correcting,
  title={Correcting direction-dependent gains in the deconvolution of radio interferometric images},
  author={Bhatnagar, S and Cornwell, TJ and Golap, K and Uson, Juan M},
  journal={Astronomy \& Astrophysics},
  volume={487},
  number={1},
  pages={419--429},
  year={2008},
  doi = {10.1051/0004-6361:20079284},
  publisher={EDP Sciences}
}

@article{tasse2018faceting,
  title={Faceting for direction-dependent spectral deconvolution},
  author={Tasse, C and Hugo, B and Mirmont, M and Smirnov, O and Atemkeng, M and Bester, L and Hardcastle, MJ and Lakhoo, R and Perkins, S and Shimwell, T},
  journal={Astronomy \& Astrophysics},
  volume={611},
  pages={A87},
  year={2018},
  publisher={EDP Sciences},
  doi={10.1051/0004-6361/201731474}
}

@ARTICLE{mitchell2008rts,
  author={Mitchell, Daniel A. and Greenhill, Lincoln J. and Wayth, Randall B. and Sault, Robert J. and Lonsdale, Colin J. and Cappallo, Roger J. and Morales, Miguel F. and Ord, Stephen M.},
  journal={IEEE Journal of Selected Topics in Signal Processing}, 
  title={Real-Time Calibration of the Murchison Widefield Array}, 
  year={2008},
  volume={2},
  number={5},
  pages={707-717},
  doi={10.1109/JSTSP.2008.2005327}}

@article{brackenhoff2025robust,
  title={Robust direction-dependent gain-calibration of beam-modelling errors far from the target field},
  author={Brackenhoff, SA and Offringa, AR and Mevius, M and Koopmans, LVE and Chege, JK and Ceccotti, E and H{\"o}fer, C and Gao, L and Ghosh, S and Mertens, FG and others},
  journal={Monthly Notices of the Royal Astronomical Society},
  volume={541},
  number={4},
  pages={3993--4010},
  year={2025},
  publisher={Oxford University Press},
  doi={10.1093/mnras/staf1206}
}

@article{kiefner2021holographic,
  title={Holographic calibration of phased array telescopes},
  author={Kiefner, U and Wayth, RB and Davidson, DB and Sokolowski, M},
  journal={Radio Science},
  volume={56},
  number={5},
  pages={e2020RS007171},
  year={2021},
  publisher={Wiley Online Library},
  doi={10.1029/2020RS007171}
}

@article{thyagarajan2025comparison,
  title={Comparison of fast, hybrid imaging architectures for multi-scale, hierarchical aperture arrays},
  author={Thyagarajan, Nithyanandan},
  journal={Publications of the Astronomical Society of Australia},
  volume={42},
  pages={e001},
  year={2025},
  publisher={Cambridge University Press},
  doi={10.1017/pasa.2024.121}
}

@article{cornwall1981selfcal,
    author = {Cornwell, T. J. and Wilkinson, P. N.},
    title = {A new method for making maps with unstable radio interferometers},
    journal = {Monthly Notices of the Royal Astronomical Society},
    volume = {196},
    number = {4},
    pages = {1067-1086},
    year = {1981},
    month = {10},
    issn = {0035-8711},
    doi = {10.1093/mnras/196.4.1067},
    url = {https://doi.org/10.1093/mnras/196.4.1067},
    eprint = {https://academic.oup.com/mnras/article-pdf/196/4/1067/3058612/mnras196-1067.pdf},
}

@article{gan2023assessing,
  title={Assessing the impact of two independent direction-dependent calibration algorithms on the LOFAR 21 cm signal power spectrum-And applications to an observation of a field flanking the north celestial pole},
  author={Gan, H and Mertens, FG and Koopmans, LVE and Offringa, AR and Mevius, M and Pandey, VN and Brackenhoff, Stefanie A and Ceccotti, E and Ciardi, B and Gehlot, BK and others},
  journal={Astronomy \& Astrophysics},
  volume={669},
  pages={A20},
  year={2023},
  publisher={EDP Sciences},
  doi={10.1051/0004-6361/202244316}
}

@article{TheHERACollaboration2025FirstII,
doi = {10.3847/1538-4357/ae2d54},
url = {https://doi.org/10.3847/1538-4357/ae2d54},
year = {2026},
month = {feb},
publisher = {The American Astronomical Society},
volume = {998},
number = {1},
pages = {33},
author = {Abdurashidova, Zuhra and Adams, Tyrone and Aguirre, James E. and Baartman, Rushelle and Barkana, Rennan and Berkhout, Lindsay M. and Bernardi, Gianni and Billings, Tashalee S. and Bizarria, Bruno B. and Bowman, Judd D. and Breitman, Daniela and Bull, Philip and Burba, Jacob and Byrne, Ruby and Carey, Steven and Chandra, Rajorshi Sushovan and Chen, Kai-Feng and Choudhuri, Samir and Cox, Tyler and DeBoer, David R. and de Lera Acedo, Eloy and Dexter, Matt and Dhandha, Jiten and Dillon, Joshua S. and Dynes, Scott and Eksteen, Nico and Ely, John and Ewall-Wice, Aaron and Fagnoni, Nicolas and Fialkov, Anastasia and Furlanetto, Steven R. and Gale-Sides, Kingsley and Garsden, Hugh and Gorce, Adelie and Gorthi, Deepthi and Halday, Ziyaad and Hazelton, Bryna J. and Hewitt, Jacqueline N. and Hickish, Jack and Huang, Tian and Jacobs, Daniel C. and Josaitis, Alec and Kern, Nicholas S. and Kerrigan, Joshua and Kittiwisit, Piyanat and Kolopanis, Matthew and Lanman, Adam and La Plante, Paul and Liu, Adrian and Ma, Yin-Zhe and MacMahon, David H. E. and Malan, Lourence and Malgas, Cresshim and Malgas, Keith and Marero, Bradley and Martinot, Zachary E. and McBride, Lisa and Mesinger, Andrei and Mirocha, Jordan and Mohamed-Hinds, Nicel and Molewa, Mathakane and Morales, Miguel F. and Muñoz, Julian B. and Murray, Steven G. and Nikolic, Bojan and Nuwegeld, Hans and Parsons, Aaron R. and Pascua, Robert and Patra, Nipanjana and Pochinda, Simon and Qin, Yuxiang and Rath, Eleanor and Razavi-Ghods, Nima and Riley, Daniel and Rosie, Kathryn and Santos, Mario G. and Singh, Saurabh and Storer, Dara and Swarts, Hilton and Tan, Jianrong and Thélie, Emilie and van Wyngaarden, Pieter and Wilensky, Michael J. and Williams, Peter K. G. and Zheng, Haoxuan and (The HERA Collaboration)},
title = {First Results from HERA Phase II},
journal = {The Astrophysical Journal}
}

@article{cumner2024effects,
  title={The effects of the antenna power pattern uncertainty within a global 21 cm experiment},
  author={Cumner, John and Pieterse, Carla and De Villiers, Dirk and de Lera Acedo, Eloy},
  journal={Monthly Notices of the Royal Astronomical Society},
  volume={531},
  number={4},
  pages={4734--4745},
  year={2024},
  publisher={Oxford University Press},
  doi={10.1093/mnras/stae1475}
}

@article{koopmans2010ionospheric,
  title={Ionospheric power-spectrum tomography in radio interferometry},
  author={Koopmans, LVE},
  journal={The Astrophysical Journal},
  volume={718},
  number={2},
  pages={963},
  year={2010},
  publisher={IOP Publishing},
  doi={10.1088/0004-637X/718/2/963}
}

@INPROCEEDINGS{anstey_2024,
  author={Anstey, Dominic and Cumner, John and Gueuning, Quentin and O'Hara, Oscar and de Lera Acedo, Eloy and Brown, Anthony and Faulkner, Andrew and Dulwich, Fred and Scott, Paul},
  booktitle={2024 18th European Conference on Antennas and Propagation (EuCAP)}, 
  title={Mitigating Zenith Blindness from Mutual Coupling in a Sunflower Phased Array}, 
  year={2024},
  volume={},
  number={},
  pages={1-5},
  doi={10.23919/EuCAP60739.2024.10501737}}

@article{bonaldi_2025,
    author = {Bonaldi, A and Hartley, P and Braun, R and Purser, S and Acharya, A and Ahn, K and Aparicio Resco, M and Bait, O and Bianco, M and Chakraborty, A and Chapman, E and Chatterjee, S and Chege, K and Chen, H and Chen, X and Chen, Z and Conaboy, L and Cruz, M and Darriba, L and De Santis, M and Denzel, P and Diao, K and Feron, J and Finlay, C and Gehlot, B and Ghosh, S and Giri, S K and Grumitt, R and Hong, S E and Ito, T and Jiang, M and Jordan, C and Kim, S and Kim, M and Kim, J and Krishna, S P and Kulkarni, A and López-Caniego, M and Labadie-García, I and Lee, H and Lee, D and Lee, N and Line, J and Liu, Y and Mao, Y and Mazumder, A and Mertens, F G and Munshi, S and Nasirudin, A and Ni, S and Nistane, V and Norregaard, C and Null, D and Offringa, A and Oh, M and Oh, S-H and Parkinson, D and Pritchard, J and Ruiz-Granda, M and Salvador López, V and Shan, H and Sharma, R and Trott, C and Yoshiura, S and Zhang, L and Zhang, X and Zheng, Q and Zhu, Z and Zuo, S and Akahori, T and Alberto, P and Allys, E and An, T and Anstey, D and Baek, J and Basavraj and Brackenhoff, S and Browne, P and Ceccotti, E and Chen, H and Chen, T and Choudhuri, S and Choudhury, M and Coles, J and Cook, J and Cornu, D and Cunnington, S and Das, S and de Lera Acedo Acedo, E and Delouis, J-M and Deng, F and Ding, J and Elahi, K M A and Fernandez, P and Fernández, C and Fernández Alcázar, A and Galluzzi, V and Gao, L-Y and Garain, U and Garrido, J and Gendron-Marsolais, M-L and Gessey-Jones, T and Ghorbel, H and Gong, Y and Guo, S and Hasegawa, K and Hayashi, T and Herranz, D and Holanda, V and Holloway, A J and Hothi, I and Höfer, C and Jelić, V and Jiang, Y and Jiang, X and Kang, H and Kim, J-Y and Koopmans, L V and Lacroix, R and Lee, E and Leeney, S and Levrier, F and Li, Y and Liu, Y and Ma, Q and Meriot, R and Mesinger, A and Mevius, M and Minoda, T and Miville-Deschênes, M-A and Moldon, J and Mondal, R and Murmu, C and Murray, S and Nirmala SR and Niu, Q and Nunhokee, C and O’Hara, O and Pal, S K and Pal, S and Park, J and Parra, M and Patra, N N and Pindor, B and Remazeilles, M and Rey, P and Rubino-Martin, J A and Saha, S and Selvaraj, A and Semelin, B and Shah, R and Shao, Y and Shaw, A K and Shi, F and Shimabukuro, H and Singh, G and Sohn, B W and Stagni, M and Starck, J-L and Sui, C and Swinbank, J D and Sánchez, J and Sánchez-Expósito, S and Takahashi, K and Takeuchi, T and Tripathi, A and Verdes-Montenegro, L and Vielva, P and Vitello, F R and Wang, G-J and Wang, Q and Wang, X and Wang, Y and Wang, Y-X and Wiegert, T and Wild, A and Williams, W L and Wolz, L and Wu, X and Wu, P and Xia, J-Q and Xu, Y and Yan, R and Yan, Y-P and Yin, Z and You, Z and Yu, X and Yu, K and Yue, B and Zhang, L and Zhao, Z and Zhao, X and Zhou, X},
    title = {Square Kilometre Array Science Data Challenge 3a: foreground removal for an EoR experiment},
    journal = {Monthly Notices of the Royal Astronomical Society},
    volume = {543},
    number = {2},
    pages = {1092-1119},
    year = {2025},
    month = {10},
    issn = {0035-8711},
    doi = {10.1093/mnras/staf1466},
    url = {https://doi.org/10.1093/mnras/staf1466},
    eprint = {https://academic.oup.com/mnras/article-pdf/543/2/1092/64222444/staf1466.pdf},
}

@ARTICLE{gueuning_2025,
  author={Gueuning, Quentin and de Lera Acedo, Eloy and Keith Brown, Anthony and Craeye, Christophe and O’Hara, Oscar},
  journal={IEEE Transactions on Antennas and Propagation}, 
  title={A Broadband Multipole Method for Accelerated Mutual Coupling Analysis of Large Irregular Arrays Including Rotated Antennas}, 
  year={2025},
  volume={73},
  number={5},
  pages={3133-3145},
  doi={10.1109/TAP.2025.3528766}}

@ARTICLE{paonessa_2023,
  author={Paonessa, Fabio and Ciorba, Lorenzo and Kyriakou, Georgios and Bolli, Pietro and Virone, Giuseppe},
  journal={IEEE Antennas and Wireless Propagation Letters}, 
  title={UAV-Based Measurement of Sharp Spectral Resonances in Mutually Coupled SKA-Low Elements}, 
  year={2023},
  volume={22},
  number={11},
  pages={2735-2739},
  doi={10.1109/LAWP.2023.3290967}}

@INPROCEEDINGS{davidson_2020,
  author={Davidson, David B. and Bolli, Pietro and Bercigli, Mirko and Ninni, Paola Di and Monari, Jader and Perini, Federico and Sokolowski, Marcin and Tingay, Steven and Ung, Daniel and Virone, Giuseppe and Waterson, Mark and Wayth, Randall and Zerbi, Filippo M.},
  booktitle={2020 XXXIIIrd General Assembly and Scientific Symposium of the International Union of Radio Science}, 
  title={Electromagnetic modelling of the SKA-LOW AAVS2 prototype}, 
  year={2020},
  volume={},
  number={},
  pages={1-4},
  doi={10.23919/URSIGASS49373.2020.9232307}}

@article{bolli_2022,
author = {Pietro Bolli and David B. Davidson and Mirko Bercigli and Paola Di Ninni and Maria Grazia Labate and Daniel Ung and Giuseppe Virone},
title = {{Computational electromagnetics for the SKA-Low prototype station AAVS2}},
volume = {8},
journal = {Journal of Astronomical Telescopes, Instruments, and Systems},
number = {1},
publisher = {SPIE},
pages = {011017},
year = {2022},
doi = {10.1117/1.JATIS.8.1.011017}
}

@article{amiri2024holographic,
  title={Holographic beam measurements of the Canadian hydrogen intensity mapping experiment (CHIME)},
  author={Amiri, Mandana and Chakraborty, Arnab and Foreman, Simon and Halpern, Mark and Hill, Alex S and Hinshaw, Gary and Landecker, TL and MacEachern, Joshua and Masui, Kiyoshi W and Mena-Parra, Juan and others},
  journal={The Astrophysical Journal},
  volume={976},
  number={2},
  pages={163},
  year={2024},
  doi = {10.3847/1538-4357/ad8133},
  publisher={The American Astronomical Society}
}

@article{chokshi_2021,
    author = {Chokshi, A and Line, J L B and Barry, N and Ung, D and Kenney, D and McPhail, A and Williams, A and Webster, R L},
    title = {Dual polarization measurements of MWA beampatterns at 137 MHz},
    journal = {Monthly Notices of the Royal Astronomical Society},
    volume = {502},
    number = {2},
    pages = {1990-2004},
    year = {2021},
    month = {04},
    issn = {0035-8711},
    doi = {10.1093/mnras/stab156},
    url = {https://doi.org/10.1093/mnras/stab156},
    eprint = {https://academic.oup.com/mnras/article-pdf/502/2/1990/36204735/stab156.pdf},
}

@article{ohara_under,
    author = {O’Hara, Oscar Sage David and Dulwich, Fred and de Lera Acedo, Eloy and Dhandha, Jiten and Gessey-Jones, Thomas and Anstey, Dominic and Fialkov, Anastasia},
    title = {Understanding spectral artefacts in SKA-Low 21-cm cosmology experiments: the impact of cable reflections},
    journal = {Monthly Notices of the Royal Astronomical Society},
    volume = {533},
    number = {3},
    pages = {2876-2892},
    year = {2024},
    month = {09},
    issn = {0035-8711},
    doi = {10.1093/mnras/stae1952},
    url = {https://doi.org/10.1093/mnras/stae1952},
    eprint = {https://academic.oup.com/mnras/article-pdf/533/3/2876/58975943/stae1952.pdf},
}

@INPROCEEDINGS{acedo_skala4,
  author={de Lera Acedo, Eloy and Pienaar, Hardie and Fagnoni, Nicolas},
  booktitle={2018 International Conference on Electromagnetics in Advanced Applications (ICEAA)}, 
  title={Antenna design for the SKA1-LOW and HERA super radio telescopes}, 
  year={2018},
  volume={},
  number={},
  pages={636-639},
  doi={10.1109/ICEAA.2018.8520395}}

@article{liu_gpr,
    author = {Liu, Yuchen and de Lera Acedo, Eloy and Sims, Peter H},
    title = {Bayesian model comparison and validation with Gaussian Process Regression for interferometric 21-cm signal recovery},
    journal = {Monthly Notices of the Royal Astronomical Society},
    volume = {549},
    number = {1},
    pages = {stag878},
    year = {2026},
    month = {06},
    issn = {0035-8711},
    doi = {10.1093/mnras/stag878},
    url = {https://doi.org/10.1093/mnras/stag878},
    eprint = {https://academic.oup.com/mnras/article-pdf/549/1/stag878/68256924/stag878.pdf},
}

@article{mertens_gpr,
    author = {Mertens, F G and Mevius, M and Koopmans, L V E and Offringa, A R and Mellema, G and Zaroubi, S and Brentjens, M A and Gan, H and Gehlot, B K and Pandey, V N and Sardarabadi, A M and Vedantham, H K and Yatawatta, S and Asad, K M B and Ciardi, B and Chapman, E and Gazagnes, S and Ghara, R and Ghosh, A and Giri, S K and Iliev, I T and Jelić, V and Kooistra, R and Mondal, R and Schaye, J and Silva, M B},
    title = {Improved upper limits on the 21 cm signal power spectrum of neutral hydrogen at z ≈ 9.1 from LOFAR},
    journal = {Monthly Notices of the Royal Astronomical Society},
    volume = {493},
    number = {2},
    pages = {1662-1685},
    year = {2020},
    month = {04},
    issn = {0035-8711},
    doi = {10.1093/mnras/staa327},
    url = {https://doi.org/10.1093/mnras/staa327},
    eprint = {https://academic.oup.com/mnras/article-pdf/493/2/1662/32666766/staa327.pdf},
}

@article{morales_diversity,
    author = {Morales, Miguel F and Beardsley, Adam and Pober, Jonathan and Barry, Nichole and Hazelton, Bryna and Jacobs, Daniel and Sullivan, Ian},
    title = {Understanding the diversity of 21 cm cosmology analyses},
    journal = {Monthly Notices of the Royal Astronomical Society},
    volume = {483},
    number = {2},
    pages = {2207-2216},
    year = {2019},
    month = {02},
    issn = {0035-8711},
    doi = {10.1093/mnras/sty2844},
    url = {https://doi.org/10.1093/mnras/sty2844},
    eprint = {https://academic.oup.com/mnras/article-pdf/483/2/2207/27159141/sty2844.pdf},
}

@misc{oskar,
  author       = {Dulwich, Fred},
  title        = {OSKAR 2.7.6},
  month        = jan,
  year         = 2020,
  note         = {{https://github.com/OxfordSKA/OSKAR/releases/tag/2. 
                   7.6}},
  publisher    = {Zenodo},
  version      = {2.7.6},
  doi          = {10.5281/zenodo.3758491}
}

@ARTICLE{ha_2018,
  author={Bui-Van, Ha and Abraham, Jens and Arts, Michel and Gueuning, Quentin and Raucy, Christopher and González-Ovejero, David and de Lera Acedo, Eloy and Craeye, Christophe},
  journal={IEEE Transactions on Antennas and Propagation}, 
  title={Fast and Accurate Simulation Technique for Large Irregular Arrays}, 
  year={2018},
  volume={66},
  number={4},
  pages={1805-1817},
  doi={10.1109/TAP.2018.2806222}}

@article{parsons2014new,
  title={New limits on 21 cm epoch of reionization from paper-32 consistent with an x-ray heated intergalactic medium at z= 7.7},
  author={Parsons, Aaron R and Liu, Adrian and Aguirre, James E and Ali, Zaki S and Bradley, Richard F and Carilli, Chris L and DeBoer, David R and Dexter, Matthew R and Gugliucci, Nicole E and Jacobs, Daniel C and others},
  journal={The Astrophysical Journal},
  volume={788},
  number={2},
  pages={106},
  year={2014},
  publisher={The American Astronomical Society},
  doi={10.1088/0004-637X/788/2/106}
}

@INPROCEEDINGS{cumner_dipoles,
  author={Cumner, John and Cumner, John and Anstey, Dominic and Gueuning, Quentin and O’Hara, Oscar and Acedo, Eloy de Lera and Brown, Anthony and Faulkner, Andrew and Dulwich, Fred and Scott, Paul},
  booktitle={2024 International Conference on Electromagnetics in Advanced Applications (ICEAA)}, 
  title={Antenna gain pattern blindness due to mutual coupling in broadband arrays}, 
  year={2024},
  volume={},
  number={},
  pages={171-171},
  doi={10.1109/ICEAA61917.2024.10702000}}

@article{mesinger_21cmfast,
    author = {Mesinger, Andrei and Furlanetto, Steven and Cen, Renyue},
    title = {21cmfast: a fast, seminumerical simulation of the high-redshift 21-cm signal},
    journal = {Monthly Notices of the Royal Astronomical Society},
    volume = {411},
    number = {2},
    pages = {955-972},
    year = {2011},
    month = {02},
    issn = {0035-8711},
    doi = {10.1111/j.1365-2966.2010.17731.x},
    url = {https://doi.org/10.1111/j.1365-2966.2010.17731.x},
    eprint = {https://academic.oup.com/mnras/article-pdf/411/2/955/4099991/mnras0411-0955.pdf},
}

@article{hurley_gleam,
    author = {Hurley-Walker, N and Callingham, J R and Hancock, P J and Franzen, T M O and Hindson, L and Kapińska, A D and Morgan, J and Offringa, A R and Wayth, R B and Wu, C and Zheng, Q and Murphy, T and Bell, M E and Dwarakanath, K S and For, B and Gaensler, B M and Johnston-Hollitt, M and Lenc, E and Procopio, P and Staveley-Smith, L and Ekers, R and Bowman, J D and Briggs, F and Cappallo, R J and Deshpande, A A and Greenhill, L and Hazelton, B J and Kaplan, D L and Lonsdale, C J and McWhirter, S R and Mitchell, D A and Morales, M F and Morgan, E and Oberoi, D and Ord, S M and Prabu, T and Shankar, N Udaya and Srivani, K S and Subrahmanyan, R and Tingay, S J and Webster, R L and Williams, A and Williams, C L},
    title = {GaLactic and Extragalactic All-sky Murchison Widefield Array (GLEAM) survey – I. A low-frequency extragalactic catalogue},
    journal = {Monthly Notices of the Royal Astronomical Society},
    volume = {464},
    number = {1},
    pages = {1146-1167},
    year = {2017},
    month = {01},
    issn = {0035-8711},
    doi = {10.1093/mnras/stw2337},
    url = {https://doi.org/10.1093/mnras/stw2337},
    eprint = {https://academic.oup.com/mnras/article-pdf/464/1/1146/47736764/mnras_464_1_1146.pdf},
}

@article{zheng_gsm,
    author = {Zheng, H. and Tegmark, M. and Dillon, J. S. and Kim, D. A. and Liu, A. and Neben, A. R. and Jonas, J. and Reich, P. and Reich, W.},
    title = {An improved model of diffuse galactic radio emission from 10 MHz to 5 THz},
    journal = {Monthly Notices of the Royal Astronomical Society},
    volume = {464},
    number = {3},
    pages = {3486-3497},
    year = {2017},
    month = {01},
    issn = {0035-8711},
    doi = {10.1093/mnras/stw2525},
    url = {https://doi.org/10.1093/mnras/stw2525},
    eprint = {https://academic.oup.com/mnras/article-pdf/464/3/3486/18518941/stw2525.pdf},
}

@article{morales2004toward,
  title={Toward epoch of reionization measurements with wide-field radio observations},
  author={Morales, Miguel F and Hewitt, Jacqueline},
  journal={The Astrophysical Journal},
  volume={615},
  number={1},
  pages={7--18},
  year={2004},
  doi={10.1086/424437}
}

@misc{skao_book_2026,
      title={Observations of the Cosmic Dawn and Epoch of Reionization with the SKAO: Observational Lessons Learned from Precursors and Pathfinder Instruments}, 
      author={Eloy de Lera Acedo and James Aguirre and Dominic Anstey and Nichole Barry and Gianni Bernardi and Somnath Bharadwaj and Anthony Brown and Jean Cavillot and Suman Chatterjee and Samir Choudhuri and Tyler Cox and John Cumner and Abhirup Datta and Fred Dulwich and Khandakar Md Asif Elahi and Andrew Faulkner and Sukhdeep Singh Gill and Quentin Gueuning and Daniel C. Jacobs and Nicholas Kern and Piyanat Kittiwisit and Yuchen Liu and Zachary Martinot and Ashish Mhaske and Florent Mertens and Vincent McKay and Satyapan Munshi and Steven Murray and Chuneeta D. Nunhokee and Oscar Sage David O'Hara and Samit K. Pal and Robert Pascua and Rashmi Sagar and Shouvik Sarkar and Shiv Sethi and Sarod Yatawatta and Oskar Zetterstrom},
      year={2026},
      eprint={2606.26072},
      archivePrefix={arXiv},
      primaryClass={astro-ph.IM},
      url={https://arxiv.org/abs/2606.26072}, 
}

@INPROCEEDINGS{ohara_impact,
  author={O’Hara, Oscar Sage David and Gueuning, Quentin and De Lera Acedo, Eloy and Dulwich, Fred and Cumner, John and Anstey, Dominic and Brown, Anthony and Faulkner, Andrew and Horton, Maya and Mhaske, Ashish},
  booktitle={2025 URSI Asia-Pacific Radio Science Meeting (AP-RASC)}, 
  title={Impact of Array Mutual Coupling on 21-cm Cosmology Experiments with the SKA-Low Telescope}, 
  year={2025},
  volume={},
  number={},
  pages={1-4},
  doi={10.46620/URSIAPRASC25/EBKD8079}}

@article{kern2019mitigating,
  title={Mitigating internal instrument coupling for 21 cm cosmology. I. Temporal and spectral modeling in simulations},
  author={Kern, Nicholas S and Parsons, Aaron R and Dillon, Joshua S and Lanman, Adam E and Fagnoni, Nicolas and de Lera Acedo, Eloy},
  journal={The Astrophysical Journal},
  volume={884},
  number={2},
  pages={105},
  year={2019},
  publisher={The American Astronomical Society},
  doi={10.3847/1538-4357/ab3e73}
}

@article{jones1941new,
  title={A new calculus for the treatment of optical systems I. Description and discussion of the calculus},
  author={Jones, R Clark},
  journal={Journal of the Optical Society of America},
  volume={31},
  number={7},
  pages={488--493},
  year={1941},
  publisher={Optical Society of America},
  doi={10.1364/JOSA.31.000488}
}

@article{rath_2025,
    author = {Rath, E and Pascua, R and Josaitis, A T and Ewall-Wice, A and Fagnoni, N and de Lera Acedo, E and Martinot, Z E and Abdurashidova, Z and Adams, T and Aguirre, J E and Baartman, R and Beardsley, A P and Berkhout, L M and Bernardi, G and Billings, T S and Bowman, J D and Bull, P and Burba, J and Byrne, R and Carey, S and Chen, K -F and Choudhuri, S and Cox, T and DeBoer, D R and Dexter, M and Dillon, J S and Dynes, S and Eksteen, N and Ely, J and Fritz, R and Furlanetto, S R and Gale-Sides, K and Garsden, H and Gehlot, B K and Ghosh, A and Gorce, A and Gorthi, D and Halday, Z and Hazelton, B J and Hewitt, J N and Hickish, J and Huang, T and Jacobs, D C and Kern, N S and Kerrigan, J and Kittiwisit, P and Kolopanis, M and Lanman, A and Liu, A and Ma, Y -Z and MacMahon, D H E and Malan, L and Malgas, C and Malgas, K and Marero, B and McBride, L and Mesinger, A and Mohamed-Hinds, N and Molewa, M and Morales, M F and Murray, S G and Nikolic, B and Nuwegeld, H and Parsons, A R and Patra, N and Plante, P La and Qin, Y and Razavi-Ghods, N and Riley, D and Robnett, J and Rosie, K and Santos, M G and Sims, P and Singh, S and Storer, D and Swarts, H and Tan, J and Wilensky, M J and Williams, P K G and van Wyngaarden, P and Zheng, H},
    title = {Investigating mutual coupling in the hydrogen epoch of reionization array and mitigating its effects on the 21-cm power spectrum},
    journal = {Monthly Notices of the Royal Astronomical Society},
    volume = {541},
    number = {2},
    pages = {1125-1144},
    year = {2025},
    month = {08},
    issn = {0035-8711},
    doi = {10.1093/mnras/staf1012},
    url = {https://doi.org/10.1093/mnras/staf1012},
    eprint = {https://academic.oup.com/mnras/article-pdf/541/2/1125/63607846/staf1012.pdf},
}

@article{liu_redundant_2010,
    author = {Liu, Adrian and Tegmark, Max and Morrison, Scott and Lutomirski, Andrew and Zaldarriaga, Matias},
    title = {Precision calibration of radio interferometers using redundant baselines},
    journal = {Monthly Notices of the Royal Astronomical Society},
    volume = {408},
    number = {2},
    pages = {1029-1050},
    year = {2010},
    month = {10},
    issn = {0035-8711},
    doi = {10.1111/j.1365-2966.2010.17174.x},
    url = {https://doi.org/10.1111/j.1365-2966.2010.17174.x},
    eprint = {https://academic.oup.com/mnras/article-pdf/408/2/1029/18440559/mnras0408-1029.pdf},
}

@article{thyagarajan_epic,
    author = {Thyagarajan, Nithyanandan and Beardsley, Adam P. and Bowman, Judd D. and Morales, Miguel F.},
    title = {A generic and efficient E-field Parallel Imaging Correlator for next-generation radio telescopes},
    journal = {Monthly Notices of the Royal Astronomical Society},
    volume = {467},
    number = {1},
    pages = {715-730},
    year = {2017},
    month = {05},
    issn = {0035-8711},
    doi = {10.1093/mnras/stx113},
    url = {https://doi.org/10.1093/mnras/stx113},
    eprint = {https://academic.oup.com/mnras/article-pdf/467/1/715/10493696/stx113.pdf},
}

@article{tingay_mwa, 
    title={The Murchison Widefield Array: The Square Kilometre Array Precursor at Low Radio Frequencies}, 
    volume={30}, 
    DOI={10.1017/pasa.2012.007}, 
    journal={Publications of the Astronomical Society of Australia}, 
    author={Tingay, S. J. and Goeke, R. and Bowman, J. D. and Emrich, D. and Ord, S. M. and Mitchell, D. A. and Morales, M. F. and Booler, T. and Crosse, B. and Wayth, R. B. and et al.}, 
    year={2013}, 
    pages={e007}
}

@INPROCEEDINGS{zarka_nenufar,
  author={Zarka, Philippe and Coffre, Andrée and Denis, Laurent and Dumez-Viou, Cédric and Girard, Julien and Grießmeier, Jean-Mathias and Loh, Alan and Tagger, Michel},
  booktitle={2018 2nd URSI Atlantic Radio Science Meeting (AT-RASC)}, 
  title={The Low-Frequency Radiotelescope NenuFAR}, 
  year={2018},
  volume={},
  number={},
  pages={1-1},
  doi={10.23919/URSI-AT-RASC.2018.8471648}
}

@ARTICLE{mkay_sky_model_correction,
       author = {{McKay}, Luke and {Subrahmanyan}, Ravi and {Chippendale}, Aaron P. and {Bolli}, Pietro and {Kyriakou}, Georgios and {Dunning}, Alex and {Ekers}, Ronald},
        title = "{Precise measurement of the absolute sky brightness at 60-350 MHz}",
      journal = {Nature Astronomy},
         year = 2026,
        month = jun,
          doi = {10.1038/s41550-026-02888-5},
archivePrefix = {arXiv},
       eprint = {2509.11846},
 primaryClass = {astro-ph.CO},
       adsurl = {https://ui.adsabs.harvard.edu/abs/2026NatAs.tmp..122M}
}

@INPROCEEDINGS{ohara_modelling,
  author={O’Hara, Oscar Sage David and Gueuning, Quentin and de Lera Acedo, Eloy and Dulwich, Fred and Anstey, Dominic and Cumner, John and Brown, Anthony and Faulkner, Andrew and Liu, Yuchen},
  booktitle={2025 19th European Conference on Antennas and Propagation (EuCAP)}, 
  title={Modelling Large Radio Telescope Visibilities Including Array Mutual Coupling Effects}, 
  year={2025},
  volume={},
  number={},
  pages={1-5},
  doi={10.23919/EuCAP63536.2025.10999651}}

@article{jishnu_2024,
  title={Calibration of an SKA‐Low Prototype Station Using Holographic Techniques},
  author={Jishnu N. Thekkeppattu and Randall Bruce Wayth and Marcin Sokołowski},
  journal={Radio Science},
  year={2024},
  volume={59},
  url={https://api.semanticscholar.org/CorpusID:266795891},
  doi={10.1029/2023RS007847}
}

@ARTICLE{conradie_2023,
  author={Conradie, André S. and Chose, Matthews and Cilliers, Pierre I. and Botha, Matthys M.},
  journal={IEEE Transactions on Antennas and Propagation}, 
  title={Antenna Array Analysis by Iterative DGFM-Based Local Solutions}, 
  year={2023},
  volume={71},
  number={6},
  pages={5199-5211},
  doi={10.1109/TAP.2023.3268726}}

@article{labate2022highlights,
  title={Highlights of the square kilometre array low frequency (SKA-LOW) telescope},
  author={Labate, Maria G and Waterson, Mark and Alachkar, Bassem and Hendre, Aniket and Lewis, Peter and Bartolini, Marco and Dewdney, Peter},
  journal={Journal of Astronomical Telescopes, Instruments, and Systems},
  volume={8},
  number={1},
  pages={011024--011024},
  year={2022},
  publisher={Society of Photo-Optical Instrumentation Engineers},
  doi={10.1117/1.JATIS.8.1.011024}
}

@article{van2018image,
  title={Image Domain Gridding: a fast method for convolutional resampling of visibilities},
  author={Van der Tol, Sebastiaan and Veenboer, Bram and Offringa, Andr{\'e} R},
  journal={Astronomy \& Astrophysics},
  volume={616},
  pages={A27},
  year={2018},
  publisher={EDP Sciences},
  doi={10.1051/0004-6361/201832858}
}

@article{comoretto2017signal,
  title={The signal processing firmware for the low frequency aperture array},
  author={Comoretto, Gianni and Chiello, Riccardo and Roberts, Matt and Halsall, Rob and Adami, Kristian Zarb and Alderighi, Monica and Aminaei, Amin and Baker, Jeremy and Belli, Carolina and Chiarucci, Simone and others},
  journal={Journal of Astronomical Instrumentation},
  volume={6},
  number={01},
  pages={1641015},
  year={2017},
  publisher={World Scientific},
  doi={10.1142/S2251171716410154}
}

@article{taylor1955design,
  title={Design of line-source antennas for narrow beamwidth and low side lobes},
  author={Taylor, Thomas Tallott},
  journal={Transactions of the IRE Professional Group on Antennas and Propagation},
  volume={3},
  number={1},
  pages={16--28},
  year={1955},
  publisher={IEEE},
  doi={10.1109/TPGAP.1955.5720407}
}

@article{mort2016analysing,
  title={Analysing the impact of far-out side-lobes on the imaging performance of the SKA-LOW telescope},
  author={Mort, Benjamin and Dulwich, Fred and Razavi-Ghods, Nima and de Lera Acedo, Eloy and Grainge, Keith},
  journal={Monthly Notices of the Royal Astronomical Society},
  pages={stw2814},
  year={2016},
  publisher={Oxford University Press},
  doi={10.1093/mnras/stw2814}
}

@article{steyskal1990mutual,
  title={Mutual coupling compensation in small array antennas},
  author={Steyskal, Hans and Herd, Jeffrey S},
  journal={IEEE Transactions on Antennas and Propagation},
  volume={38},
  number={12},
  pages={1971--1975},
  year={1990},
  publisher={IEEE},
  doi={10.1109/8.60990}
}

@article{macario2022characterization,
  title={Characterization of the SKA1-low prototype station aperture array verification System 2},
  author={Macario, Giulia and Pupillo, Giuseppe and Bernardi, Gianni and Bolli, Pietro and Di Ninni, Paola and Comoretto, Giovanni and Mattana, Andrea and Monari, Jader and Perini, Federico and Schiaffino, Marco and others},
  journal={Journal of Astronomical Telescopes, Instruments, and Systems},
  volume={8},
  number={1},
  pages={011014--011014},
  year={2022},
  publisher={Society of Photo-Optical Instrumentation Engineers},
  doi={10.1117/1.JATIS.8.1.011014}
}

@article{benthem2021aperture,
  title={The aperture array verification system 1: system overview and early commissioning results},
  author={Benthem, P and Wayth, R and de Lera Acedo, E and Zarb Adami, Kristian and Alderighi, MONICA and Belli, Carolina and Bolli, Pietro and Booler, T and Borg, Josef and Broderick, JW and others},
  journal={Astronomy \& Astrophysics},
  volume={655},
  pages={A5},
  year={2021},
  publisher={EDP Sciences},
  doi={10.1051/0004-6361/202040086}
}

@article{joseph2020calibration,
  title={Calibration and 21-cm power spectrum estimation in the presence of antenna beam variations},
  author={Joseph, Ronniy C and Trott, CM and Wayth, RB and Nasirudin, A},
  journal={Monthly Notices of the Royal Astronomical Society},
  volume={492},
  number={2},
  pages={2017--2028},
  year={2020},
  publisher={Oxford University Press},
  doi={10.1093/mnras/stz3375}
}

@INPROCEEDINGS{wijnholds_holography_2021,
  author={Wijnholds, Stefan J.},
  booktitle={2021 XXXIVth General Assembly and Scientific Symposium of the International Union of Radio Science (URSI GASS)}, 
  title={Generalised Self-Holography}, 
  year={2021},
  volume={},
  number={},
  pages={1-4},
  doi={10.23919/URSIGASS51995.2021.9560415}}

@article{ward1995theory,
  title={Theory and design of broadband sensor arrays with frequency invariant far-field beam patterns},
  author={Ward, Darren B and Kennedy, Rodney A and Williamson, Robert C},
  journal={The Journal of the Acoustical Society of America},
  volume={97},
  number={2},
  pages={1023--1034},
  year={1995},
  publisher={Acoustical Society of America},
  doi={10.1121/1.412215}
}

@INPROCEEDINGS{kyriakou_cma,
  author={Kyriakou, Georgios and Bolli, Pietro and Virone, Giuseppe},
  booktitle={2023 17th European Conference on Antennas and Propagation (EuCAP)}, 
  title={Characteristic Modes Analysis of Mutually Coupled Log-Periodic Dipole Antennas}, 
  year={2023},
  volume={},
  number={},
  pages={1-4},
  doi={10.23919/EuCAP57121.2023.10133524}}

@article{planck_2018,
    author = "Aghanim, N. and others",
    collaboration = "Planck",
    title = "{Planck 2018 results. VI. Cosmological parameters}",
    eprint = "1807.06209",
    archivePrefix = "arXiv",
    primaryClass = "astro-ph.CO",
    doi = "10.1051/0004-6361/201833910",
    journal = "Astron. Astrophys.",
    volume = "641",
    pages = "A6",
    year = "2020",
    note = "[Erratum: Astron.Astrophys. 652, C4 (2021)]"
}

@article{hogg1999distance,
  title={Distance measures in cosmology},
  author={Hogg, David W},
  journal={arXiv preprint astro-ph/9905116},
  year={1999}
}

@article{lanman_2020,
    author = {Lanman, Adam E and Pober, Jonathan C and Kern, Nicholas S and de Lera Acedo, Eloy and DeBoer, David R and Fagnoni, Nicolas},
    title = {Quantifying EoR delay spectrum contamination from diffuse radio emission},
    journal = {Monthly Notices of the Royal Astronomical Society},
    volume = {494},
    number = {3},
    pages = {3712-3727},
    year = {2020},
    month = {05},
    issn = {0035-8711},
    doi = {10.1093/mnras/staa987},
    url = {https://doi.org/10.1093/mnras/staa987},
    eprint = {https://academic.oup.com/mnras/article-pdf/494/3/3712/33145162/staa987.pdf},
}

@article{leshem2000multichannel,
  title={Multichannel interference mitigation techniques in radio astronomy},
  author={Leshem, Amir and Van der Veen, Alle-Jan and Boonstra, Albert-Jan},
  journal={The Astrophysical Journal Supplement Series},
  volume={131},
  number={1},
  pages={355--373},
  year={2000},
  doi={10.1086/317360}
}

@inproceedings{thyagarajan2026fast,
  title={A fast, wide-field, and real-time imaging prototype for large-aperture arrays},
  author={Thyagarajan, Nithyanandan and Thekkeppattu, Jishnu N and Humphrey, David},
  booktitle={Radio Telescopes, Technologies, and Methods},
  volume={14153},
  pages={108--116},
  year={2026},
  organization={SPIE},
  doi={10.1117/12.3105286}
}



\appendix

\section{On the Connection Between Beamforming and Interferometric Imaging}
\label{app:connection}

This appendix derives the equivalence between beamforming and interferometric imaging that underlies direct-imaging architectures such as EPIC \citep{thyagarajan_epic, thyagarajan2026fast} and the visibility-domain correction of Appendix~\ref{app:station_vis}. It works within a single station, using the steering matrices $\mathbfcal{J}_p$, beamforming weights $\mathbf{w}_l$ and brightness matrix $\mathbf{I}$ of Section~\ref{sec:modelling}. Whereas $m$ and $n$ label stations in Section~\ref{sec:modelling}, here they label antennas within the station, and $\mathbf{b}_{mn} = \mathbf{r}_m - \mathbf{r}_n$ is the baseline between them.
 
For a plane wave from direction $\hat{\mathbf{s}}$, the beamformed voltage of feed $p$ for beam $l$ (Figure~\ref{fig:beamformed}, Beamformers) is the array pattern of Equation~\eqref{eq:array_pattern} applied to the incident field,
\begin{equation}
    b_{l,p}(f,\hat{\mathbf{s}},t) = \mathbf{w}_l\, \mathbfcal{J}_p \, \mathbf{e}_\mathrm{i} = \mathbf{A}_{l,p}\, \mathbf{e}_\mathrm{i}.
    \label{eq:beamformed_voltage}
\end{equation}
For a continuous distribution of randomly time-varying far-field sources, the beamformed voltage is obtained by integrating this response over the visible hemisphere,
\begin{equation}
\label{eq:beamforming_multisource}
b_{l,p}(f,t) = \mathbf{w}_l 
\int 
\mathbfcal{J}_p \mathbf{e}_\mathrm{i}\,
\mathrm{d}\Omega,
\end{equation}
and the correlation between the beamformed voltages of feeds $p$ and $q$ is
\begin{equation}
\label{cross_corrbeams}
P_{l,pq}(f,t) = \langle b_{l,p} \, b_{l,q}^\dagger \rangle.
\end{equation}
 
Inserting Equation~\eqref{eq:beamforming_multisource} into Equation~\eqref{cross_corrbeams} and exchanging the order of integration, the beamformed cross-power can be written as
\begin{align}
P_{l,pq}(f,t) & = \mathbf{w}_l \mathbf{R}_{pq} \mathbf{w}_l^{\dagger},
\label{equiv1}
\end{align}
where $\mathbf{R}_{pq}(f, t) \in \mathbb{C}^{N_a \times N_a}$ is the array correlation matrix that captures the spatial coherence between antennas. For a spatially incoherent sky, i.e., sources whose electric fields are spatially uncorrelated, its $(m,n)$-th element, the \textit{visibility} on the baseline $\mathbf{b}_{mn}$ between antennas $m$ and $n$, is given by
\begingroup
\small
\begin{align}
    R_{pq,mn}(f, t) = \int \, \mathbf{j}_{m,p}(f,\hat{\mathbf{s}}) \, \mathbf{I}(f,\hat{\mathbf{s}},t) \, \mathbf{j}_{n,q}^{\dagger}(f,\hat{\mathbf{s}}) \, e^{\mathrm{j} k_0 \hat{\mathbf{s}} \cdot \mathbf{b}_{mn}} \, \mathrm{d}\Omega.
\end{align}
\endgroup
This has the same form as the station visibility of Equation~\eqref{eq:visibility_rime}, formed at the Correlators in Figure~\ref{fig:beamformed}, with the Jones-matrix rows $\mathbf{j}_{m,p}$ and $\mathbf{j}_{n,q}$ of individual antennas in place of the array patterns of whole stations.
 
As illustrated in the first block of Figure~\ref{fig:beamformed}~(Beam Pointings), we now extend this formulation to the case where the station forms multiple simultaneous beams toward a discrete set of $N_{\mathrm{dir}}$ steering directions $\hat{\mathbf{s}}_l$ across the $(l_x,m_y)$ sky grid. Stacking the weight vectors $\mathbf{w}_l$ of these beams as rows gives the matrix $\mathbf{W}(f) \in \mathbb{C}^{N_{\mathrm{dir}} \times N_a}$, whose $(l,n)$-th element $w_{l,n}$ is the phase compensation applied to antenna $n$ when forming a beam toward $\hat{\mathbf{s}}_l$. In this case, the beamformed power maps the sky intensity across all directions, and the $N_{\mathrm{dir}} \times N_{\mathrm{dir}}$ beam power matrix can be written as
\begin{align}
\label{eq:beam_power_matrix}
    \mathbf{P}_{pq}(f,t) = \mathbf{W} \cdot \mathbf{R}_{pq} \cdot \mathbf{W}^{\dagger}.
\end{align}
This matrix $\mathbf{P}_{pq}(f,t)$ contains both the power measured in each scanned direction (its diagonal elements) and the cross-correlations between beams pointing in different directions (its off-diagonal elements). While the off-diagonal terms underpin holographic calibration techniques \citep{kiefner2021holographic}, this work focuses exclusively on the diagonal elements, which represent the power received by each formed beam. These elements define the \textit{dirty image vector},
\begin{equation}
    \mathbf{i}_{pq}(f,t) \equiv \mathrm{diag}\big(\mathbf{P}_{pq}(f,t)\big).
\end{equation}
 
To express the diagonal entries compactly, we use the identity
\begin{equation}
    \mathrm{diag}(\mathbf{A X B}) = (\mathbf{B}^{\mathrm{T}} \odot \mathbf{A})\, \mathrm{vec}(\mathbf{X}).
\end{equation}
Applying this to Equation~\eqref{eq:beam_power_matrix} gives
\begin{equation}
    \mathbf{i}_{pq}(f,t) = \mathbfcal{F}(f)\, \mathrm{vec}(\mathbf{R}_{pq}(f,t)),
    \label{eq:dirtyimage}
\end{equation}
\justify{where the \textit{Fourier matrix} is defined as $\mathbfcal{F}(f) = \mathbf{W}^{\ast}(f) \odot \mathbf{W}(f)$. In this case, the beamformed dirty image $\mathbf{i}_{pq}(f,t)$ is analogous to the classical interferometric dirty image, where $\mathbfcal{F}(f)$ is the Fourier synthesis operator that typically maps visibilities to image pixels.}

\section{Station- and Visibility-Domain Correction}
\label{app:station_vis}
 
This appendix derives the station-level operator $\mathbf{C}_\mathrm{A}$ introduced at the end of Section~\ref{sec:element-level-cal} (Section~\ref{sec:beam_level_cal}), shows how it enters the visibilities (Section~\ref{sec:vis_level_cal}), and verifies both numerically (Section~\ref{sec:beam_vis_propagation}).
 
\subsection{Station-Level Pattern Correction}
\label{sec:beam_level_cal}
 
The antenna-level fit of Section~\ref{sec:element-level-cal} can equally be posed after beamforming, correcting the beams formed by each station rather than the individual antenna voltages. Forming all $N_{\mathrm{dir}}$ beams at once in each polarisation with the weight matrix $\mathbf{W}$ of Appendix~\ref{app:connection}, and evaluating the array patterns of Equation~\eqref{eq:array_pattern} over the same $M$ directions as $\mathbfcal{J}$, the model and target array patterns are
\begin{equation}
    \mathbf{A}^{\mathrm{mod}} = (\mathbf{I}_2 \otimes \mathbf{W})\,\mathbfcal{J}^{\mathrm{mod}}, \qquad
    \mathbf{A}^{\mathrm{tgt}} = (\mathbf{I}_2 \otimes \mathbf{W})\,\mathbfcal{J}^{\mathrm{tgt}},
    \label{eq:ap_pointings}
\end{equation}
where each row corresponds to one combination of beam pointing and feed, ordered as all x-feed beams followed by all y-feed beams, so that $\mathbf{A}^{\mathrm{mod}}, \mathbf{A}^{\mathrm{tgt}} \in \mathbb{C}^{2N_{\mathrm{dir}} \times 2M}$. As at the antenna level, the two are related by a linear transformation
\begin{equation}
    \mathbf{A}^{\mathrm{tgt}} \simeq \mathbf{C}_\mathrm{A}\, \mathbf{A}^{\mathrm{mod}},
    \label{eq:beam-level}
\end{equation}
where $\mathbf{C}_\mathrm{A} \in \mathbb{C}^{2N_{\mathrm{dir}} \times 2N_{\mathrm{dir}}}$ acts on the beamformed voltages at the Beamformers stage of Figure~\ref{fig:beamformed}. In the specific case where the target is formed from the isolated element patterns and the model from the embedded element patterns, it removes the mutual-coupling-induced structure from the formed beams. Minimising the weighted Frobenius norm of Section~\ref{sec:element-level-cal}, with $\mathbf{A}^{\mathrm{mod}}$ and $\mathbf{A}^{\mathrm{tgt}}$ in place of $\mathbfcal{J}^{\mathrm{mod}}$ and $\mathbfcal{J}^{\mathrm{tgt}}$ and the same directional weights $\mathbf{D}$, gives the closed-form solution
\begin{equation}
    \mathbf{C}_\mathrm{A} = \mathbf{A}^{\mathrm{tgt}}\mathbf{D} \left(\mathbf{A}^{\mathrm{mod}} \mathbf{D}\right)^{+}.
\end{equation}
 
Because every beam is formed from the same $N_a$ element patterns, $\mathbf{A}^{\mathrm{mod}}\mathbf{D}$ can have full row rank only when $N_{\mathrm{dir}} \leq N_a$, and even then the inversion can become ill-conditioned for the same reasons as at the antenna level. We therefore again apply Tikhonov regularisation,
\begin{equation}
    \mathbf{C}_\mathrm{A} = \mathbf{A}^{\mathrm{tgt}} \mathbf{D} \left( \mathbf{A}^{\mathrm{mod}} \mathbf{D} \right)^\dagger 
        \Big[ \left( \mathbf{A}^{\mathrm{mod}} \mathbf{D} \right) \left( \mathbf{A}^{\mathrm{mod}} \mathbf{D} \right)^\dagger + \lambda_{\mathrm{reg}} \mathbf{I}_{2N_{\mathrm{dir}}} \Big]^{-1}.
    \label{eq:beam_regularised}
\end{equation}

\subsection{Visibility-Domain Pattern Construction}
\label{sec:vis_level_cal}
 
The visibility-domain correction follows by applying the beamforming-correlation equivalence of Appendix~\ref{app:connection} to the two-station case of Figure~\ref{fig:beamformed}. There, Equation~\eqref{cross_corrbeams} correlates the two polarisations of a single station beam; the same construction holds when the two beamformed voltages come from different stations $m$ and $n$. Dropping the polarisation index for clarity, the correlation of $b_{l,m}$ and $b_{l,n}$ is the visibility measured on the baseline $\mathbf{b}_{mn} = \mathbf{r}_m - \mathbf{r}_n$ between the two stations, for a beam pointed towards $\hat{\mathbf{s}}_l$,
\begingroup
\small
\begin{align}
\begin{split}
    R_{l,mn}(f,t) &= \langle b_{l,m}\, b_{l,n}^\dagger \rangle \\
    &= \int \mathbf{A}_{l,m}(f,\hat{\mathbf{s}}) \, \mathbf{I}(f,\hat{\mathbf{s}}, t) \, \mathbf{A}_{l,n}^{\dagger}(f,\hat{\mathbf{s}}) \, e^{\mathrm{j} k_0 \hat{\mathbf{s}} \cdot \mathbf{b}_{mn}} \, \mathrm{d}\Omega,
\end{split}
\end{align}
\endgroup
where $\mathbf{A}_{l,m}$ and $\mathbf{A}_{l,n}$ are the array patterns of Equation~\eqref{eq:array_pattern} for stations $m$ and $n$, as in Equation~\eqref{eq:visibility_rime}.
 
Because $\mathbf{C}_\mathrm{A}$ mixes beams and feeds, its effect is expressed through the matrix $\mathbf{R}_{mn} \in \mathbb{C}^{2N_{\mathrm{dir}} \times 2N_{\mathrm{dir}}}$ of correlations between every beam and feed of station $m$ and every beam and feed of station $n$, which contains the visibilities $R_{l,mn}$ above for every beam and polarisation product. Applying $\mathbf{C}_\mathrm{A}$ to the beamformed voltages of both stations gives the corrected visibilities
\begin{align}
\begin{split}
    \mathbf{R}^{\mathrm{tgt}}_{mn} &= \mathbf{A}^{\mathrm{tgt}}_{m} \, \mathbf{I} \, \left(\mathbf{A}_{n}^{\mathrm{tgt}}\right)^\dagger \\
    &\simeq \mathbf{C}_\mathrm{A} \mathbf{A}_{m}^{\mathrm{true}} \, \mathbf{I} \, \mathbf{A}_{n}^{\mathrm{true}\dagger}  \mathbf{C}_\mathrm{A}^\dagger \\
    &\simeq \mathbf{C}_\mathrm{A} \, \mathbf{R}_{mn}^{\mathrm{true}} \, \mathbf{C}_\mathrm{A}^\dagger,
\label{eq:vis_congruence}
\end{split}
\end{align}
\justify{where $\mathbf{A}^{\mathrm{tgt}}_{m}$ collects the target array patterns of all $N_{\mathrm{dir}}$ beams and both feeds of station $m$. The second step holds because $\mathbf{C}_\mathrm{A}$ does not depend on direction and can therefore be moved inside the integral, where, by Equation~\eqref{eq:beam-level}, it maps the model array patterns onto the target ones. This holds only for emission within the directions weighted by $\mathbf{D}$, and only if the model matches the true response, $\mathbf{A}^{\mathrm{mod}} = \mathbf{A}^{\mathrm{true}}$. Even then, the mapping is exact only to the precision of the fit, which for an unregularised, Nyquist-sampled fit reaches a numerical floor of about $-125$~dBV (Figure~\ref{fig:EP_slices_sampling_and_regularisation}); emission outside the weighted directions is not mapped onto the target, as the full-sky simulations of Figure~\ref{fig:Cal_mainbeam_vs_fullsky} show.}

\subsection{From Voltages to Visibilities}
\label{sec:beam_vis_propagation}
 
\justify{We now test whether the behaviour found at the antenna level carries over to the beamformed-voltage and visibility domains. To compute the station-level correction matrix of Equation~\eqref{eq:beam_regularised}, we formed array patterns from the Nyquist-sampled radiation patterns $\mathbfcal{J}^{\mathrm{mod}}$ and $\mathbfcal{J}^{\mathrm{tgt}}$ for $N_{\mathrm{dir}} = 256$ pointing directions distributed across the visible hemisphere on a Fibonacci lattice. Increasing $N_{\mathrm{dir}}$ does not improve the system's conditioning. As noted in Section~\ref{sec:beam_level_cal}, every beam is a linear combination of the same $N_a$ element patterns, so the rows of $\mathbf{A}^{\mathrm{mod}}\mathbf{D}$ span at most $2N_a$ dimensions. Once $N_{\mathrm{dir}} > N_a$, each additional pointing adds two rows that are linear combinations of the others, and the Gram matrix $(\mathbf{A}^{\mathrm{mod}}\mathbf{D})(\mathbf{A}^{\mathrm{mod}}\mathbf{D})^\dagger$ in Equation~\eqref{eq:beam_regularised} acquires $2(N_{\mathrm{dir}} - N_a)$ zero eigenvalues that only the regularisation keeps invertible. \citet{wijnholds_holography_2021} found the same for generalised self-holography: because all beams are formed simultaneously from the same element signals, forming more beams than elements adds no information, and the conditioning is instead set by how close to orthogonal the beamforming weight vectors are.}
 
\justify{For a like-for-like comparison with the antenna-level results, we use the same field-of-view threshold as in Section~\ref{sec:mc_ant_volt_cal} to define the binary direction weights at each pointing, so that both corrections are evaluated over an identical sky region. However, because the array-pattern amplitude scales with the number of elements, a fixed regularisation term would have a much weaker relative effect at the station level. To preserve a comparable effective strength, we scale $\lambda_{\mathrm{reg}}$ by $N_a^2$, since it is added to the Gram matrix $(\mathbf{A}^{\mathrm{mod}}\mathbf{D})(\mathbf{A}^{\mathrm{mod}}\mathbf{D})^\dagger$, which scales with the square of the array-pattern amplitude, giving $\lambda_{\mathrm{reg}} = 256^2\times10^{-4}$ for the value of $10^{-4}$ adopted at the antenna level.}
 
\begin{figure}
\centering
\includegraphics[width=.95\linewidth]{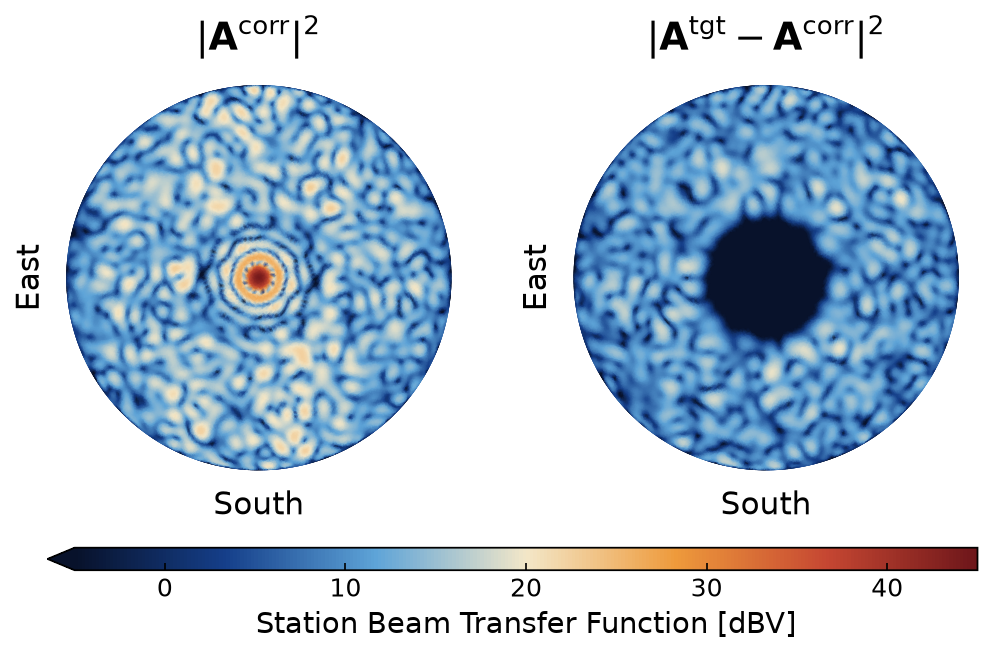}
\caption{Beamformed voltage-domain mutual-coupling compensation for the Perturbed Vogel SKA-Low station at 122~MHz, for a zenith-pointed beam. The left panel shows the corrected array pattern $|\mathbf{A}^{\mathrm{corr}}|$, while the right panel shows the residual $|\mathbf{A}^{\mathrm{tgt}} - \mathbf{A}^{\mathrm{corr}}|$ between the isolated-element and corrected array patterns. Both panels show beam transfer functions in dBV as functions of directional cosine. The correction employs Tikhonov regularisation with $\lambda_{\mathrm{reg}} = 256^2\times10^{-4}$ and is sampled at Nyquist. Within the corrected field of view, the residual is suppressed to the noise floor, while the stochastic $\sim$0--15~dBV sidelobe-level residuals outside this region mirror the uncorrected element-to-element variation seen in the station-level patterns of Figure~\ref{fig:element-level-fullsky}.}
\label{fig:beam-level-fullsky}
\end{figure}
 
\begin{figure}
    \centering
    \includegraphics[width=.95\linewidth]{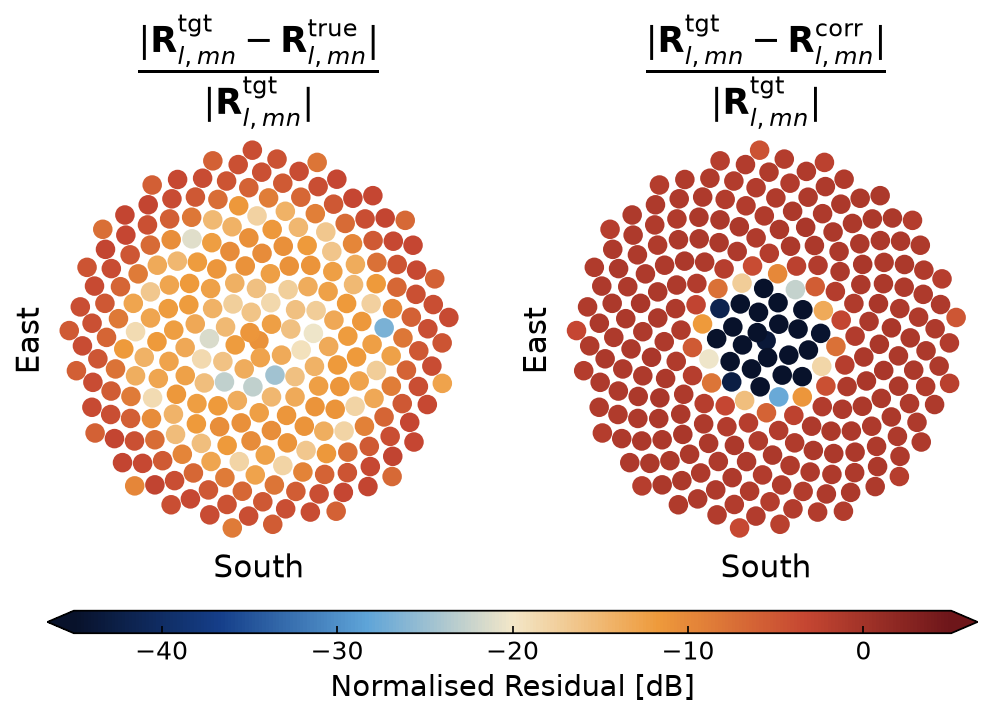}
    \caption{\justify{Visibility-domain mutual-coupling compensation for a two-station baseline of 310.8~m oriented 143.7$^\circ$ east of north at 122~MHz. Each panel displays the normalised residual of the diagonal entries of the Stokes-$I$ component of the correlation matrix, computed over all $N_{\mathrm{dir}}$ beam pointings given by each marker. The left panel shows the residual of the uncorrected visibilities, while the right panel shows the residual after application of the correction matrix.}}
    \label{fig:vis-level-fullsky}
\end{figure}
 
\justify{Figure~\ref{fig:beam-level-fullsky} shows the resulting station-level correction at 122~MHz for a zenith-pointed beam. Similar to the antenna level, the corrected array pattern (left panel) recovers a compact main lobe centred on the pointing direction, while elsewhere it retains the stochastic sidelobe structure inherited from the uncorrected EEPs, with maxima of order 20~dBV. Within the corrected field of view, the residual (right panel) falls below $-40$~dBV, below the colour scale's lower limit, showing that $\mathbf{C}_\mathrm{A}$ reproduces the isolated-element array pattern with high fidelity. Outside this region, which receives no correction from $\mathbf{C}_\mathrm{A}$, the regularised sidelobes and their residuals track the levels of the uncorrected EEPs, mirroring the array patterns in the bottom row of Figure~\ref{fig:element-level-fullsky}. Correcting directly at the station level therefore reproduces both the main-lobe restoration and the sidelobe limitations of the antenna-level correction; we return to the latter in Section~\ref{sec:mainbeam_sidelobe} when assessing the impact on EoR window recovery.}
 
\justify{We now extend this test to the visibility domain. To characterise the residuals as a function of direction without resorting to imaging, we compute visibilities via Equation~\eqref{eq:visibility_rime} for a single unit point source at each of the $N_{\mathrm{dir}}$ beam pointings, on a 310.8~m baseline oriented 143.7$^\circ$ east of north at 122~MHz. The same matrix $\mathbf{C}_\mathrm{A}$ used in Figure~\ref{fig:beam-level-fullsky} is then applied directly to the simulated visibilities $\mathbf{R}_{mn}$ through the congruence of Equation~\eqref{eq:vis_congruence}, giving the normalised residuals shown in Figure~\ref{fig:vis-level-fullsky}. Although the colour bar is saturated for clarity, the correction suppresses the residuals in the corrected pointing directions by more than 80~dB. The agreement between the station- and visibility-domain results confirms that, within the constrained directions, the correction propagates reliably to the visibility domain and, since the same matrix applies without re-computation, can be applied entirely in post-processing.}

\vspace{-.5cm}
\section{Normalisation of the Cross-Time Estimator}
\label{app:gamma}

First we write the response of the Stokes~I visibility as
\begin{equation}
    \mathcal{A}_{mn}(f,\hat{\mathbf{s}},t) = \mathrm{tr}\!\left[\mathbf{A}_{m}(f,\hat{\mathbf{s}},t)\,\mathbf{A}_{n}^{\dagger}(f,\hat{\mathbf{s}},t)\right]/2.
    \label{eq:stokes_i_response}
\end{equation} 
Here, $\mathbf{A}_m$ and $\mathbf{A}_n$ are the array patterns of Equation~\eqref{eq:array_pattern} for the pointing at time $t$, and, unlike in Section~\ref{sec:modelling}, $\hat{\mathbf{s}}$ is fixed on the sky rather than to the station, so that every sample sees the same sky and the time dependence moves from $\mathbf{I}$ to the array patterns.

\justify{If the beam impulse response is considered narrow compared to the structure of the brightness-temperature field, the per-baseline estimator approximates the power spectrum of Equation~\eqref{eq:pk_perbaseline} \citep{o2025uncovering}, where the normalisation $\Omega_\mathcal{A}B_\mathrm{eff}$ is given by the integral of $\mathcal{W}^2|\mathcal{A}_{mn}|^2$ over solid angle and frequency. Here the fringe term cancels in this integral because each sample is multiplied by its own complex conjugate. 
For the case of the cross-time estimator of Equation~\eqref{eq:pk_crosstime}, block sums $S_{\kappa,i}$ and $S_{\kappa,j}$ are instead multiplied together, thereby the same approximation holds for a single response of block $i$,}
\begin{equation}
    \mathcal{K}_{\kappa,i}(f,\hat{\mathbf{s}}) = \sum_{s\in\mathscr{S}_{\kappa,i}} \mathcal{A}_{mn}(f,\hat{\mathbf{s}},t)\, e^{\mathrm{j} k_0 (\hat{\mathbf{s}}-\hat{\mathbf{s}}_l)\cdot\mathbf{b}_{mn}},
\end{equation}
where $\mathscr{S}_{\kappa,i}$ is the set of samples $s \equiv (mn,t)$ in cell $\kappa$ and block $i$, $\hat{\mathbf{s}}_l$ is the phase centre, and the baseline $\mathbf{b}_{mn}$ is evaluated at time $t$. Because different samples are now multiplied, the fringe terms no longer cancel, and $|\mathcal{A}_{mn}|^2$ in the normalisation is replaced by the product of the two block responses,
\begin{equation}
    \Gamma_{\kappa} = \sum_{i \neq j} \Re \int \mathcal{W}^2(f) \int \mathcal{K}_{\kappa,i}(f,\hat{\mathbf{s}})\,\mathcal{K}_{\kappa,j}^{\ast}(f,\hat{\mathbf{s}})\,\mathrm{d}\Omega\,\mathrm{d}f.
    \label{eq:gamma_def}
\end{equation}
As the station tracks the field and successive blocks sample each cell through different baselines, the products $\mathcal{K}_{\kappa,i}\mathcal{K}_{\kappa,j}^{\ast}$ lose coherence with angular distance from the pointing, so $\Gamma_\kappa$ is dominated by directions near it and plays the role of $\Omega_\mathcal{A}B_\mathrm{eff}$ for the cross-time estimator. 

\bsp	
\label{lastpage}
\end{document}